\documentclass[twocolumn,showpacs,preprintnumbers,amsmath,amssymb]{revtex4-1}

\usepackage{graphicx}
\usepackage{dcolumn}
\usepackage{bm}

\usepackage{verbatim}
\usepackage{amsmath}
\usepackage{color}
\usepackage{multirow}  

\usepackage{soul}

\begin{document}

\newtheorem{theorem}{Theorem}
\newtheorem{acknowledgement}[theorem]{Acknowledgement}
\newtheorem{algorithm}[theorem]{Algorithm}
\newtheorem{axiom}[theorem]{Axiom}
\newtheorem{claim}[theorem]{Claim}
\newtheorem{conclusion}[theorem]{Conclusion}
\newtheorem{condition}[theorem]{Condition}
\newtheorem{conjecture}[theorem]{Conjecture}
\newtheorem{corollary}[theorem]{Corollary}
\newtheorem{criterion}[theorem]{Criterion}
\newtheorem{definition}[theorem]{Definition}
\newtheorem{example}[theorem]{Example}
\newtheorem{exercise}[theorem]{Exercise}
\newtheorem{lemma}[theorem]{Lemma}
\newtheorem{notation}[theorem]{Notation}
\newtheorem{problem}[theorem]{Problem}
\newtheorem{proposition}[theorem]{Proposition}
\newtheorem{remark}[theorem]{Remark}
\newtheorem{solution}[theorem]{Solution}
\newtheorem{summary}[theorem]{Summary}
\newenvironment{proof}[1][Proof]{\noindent\textbf{#1.} }{\ \rule{0.5em}{0.5em}}

\renewcommand{\theequation}{\thesection.\arabic{equation}}

\newcommand{\re}{\mathop{\mathrm{Re}}}

\newcommand{\lb}{\label}
\newcommand{\be}{\begin{equation}}
\newcommand{\ee}{\end{equation}}
\newcommand{\bea}{\begin{eqnarray}}
\newcommand{\eea}{\end{eqnarray}}


\title{Varying constants driven baryogenesis}

\author{Katarzyna Leszczy\'nska}
 \email{katarzyna.leszczynska@usz.edu.pl}
\affiliation{Institute of Physics, University of Szczecin, Wielkopolska 15, 70-451 Szczecin, Poland}%

\author{Mariusz P. D\c{a}browski}
 \email{mariusz.dabrowski@usz.edu.pl} 
 \affiliation{\it Institute of Physics, University of Szczecin, Wielkopolska 15, 70-451 Szczecin, Poland,}
\affiliation{\it National Centre for Nuclear Research, Andrzeja So{\l}tana 7, 05-400 Otwock, Poland,}
\affiliation{\it Copernicus Center for Interdisciplinary Studies,
S{\l }awkowska 17, 31-016 Krak\'ow, Poland}

\author{Tomasz Denkiewicz}
 \email{tomasz.denkiewicz@usz.edu.pl}
\affiliation{ Institute of Physics, University of Szczecin, Wielkopolska 15, 70-451 Szczecin, Poland}%
\affiliation{\it Copernicus Center for Interdisciplinary Studies,
S{\l }awkowska 17, 31-016 Krak\'ow, Poland} 

\date{\today}

\begin{abstract} 
We study the spontaneous baryogenesis scenario in the early universe for three different frameworks of varying constants theories. We replace the constants by dynamical scalar fields playing the role of thermions. We first obtain the results for baryogenesis driven by the varying gravitational constant, $G$, as in the previous literature, then challenge the problem for varying fine structure constant $\alpha$ models as well as for varying speed of light $c$ models. We show that in each of these frameworks the current observational value of the baryon to entropy ratio, $\eta_B \sim 8.6 \cdot 10^{-11}$, can be obtained for large set of parameters of dynamical constants models as well as the decoupling temperature, and the characteristic cut-off length scale. 

\end{abstract}


\maketitle


\section{\label{sec:level1}Introduction}

The problem of an excess of the matter over the antimatter in the universe we observe, is one of the biggest mysteries of contemporary cosmology. Why do we live in the particle-filled universe and not in the anti-particle-filled universe is not an obvious issue, especially taking into account that the anti-particles are observed in the particle-filled universe, too. There are series of explanations to the mystery appealing to the earliest stages of the universe evolution such as to the Planck scale quantum gravity era \cite{kusenko} or even before that, as suggested for example in the context of the multiverse concept \cite{salva,Turok2018}. 
However, despite the fact, that it is generally agreed, that quantum gravity does not preserve any global quantum numbers (as is evident from the lost of baryon number in the process of a star collapse forming a black hole), inflationary expansion is considered to dilute any such matter--antimatter asymmetry. Then, one should look for the solution of the problem in the subsequent stages of the evolution of the universe. The first attempt to explain the problem was given in the renowned paper by Sakharov \cite{sakharov}. 

As it is often referred, he suggested three necessary conditions for the matter--antimatter asymmetry to occur in the universe: the baryon number $B$ violation interactions have to appear; charge $C$ and charge-parity $CP$ violating particle processes have to be possible; departure from thermal equilibrium to shift the densities of particles with respect to antiparticles have to be present. This idea of Sakharov was developed in many ways \cite{dimopoulos78}. 

An idea of the spontaneous baryogenesis was later introduced by Cohen and Kaplan \cite{cohenkaplan}. In fact, they challenged the third Sakharov's condition, i.e. the departure from thermal equilibrium, postulating instead a spontaneous breaking of CPT symmetry already at the thermal equilibrium which generates the shift of the energy of the baryons with respect to the energy of anti-baryons in the universe which is responsible for the baryon asymmetry. An exit from thermal equilibrium takes place at some decoupling temperature, but once generated, baryon asymmetry is frozen-in in this scenario. 
Baryogenesis is driven by a scalar field---the thermion---which decays after the baryon asymmetry is established \cite{DeSimone,DeFelice}. The role of the thermion can also be played by some gravitationally motivated scalar such as Ricci or Gauss-Bonnet scalars and their combinations---such scenarios are called gravitational baryogeneses \cite{Gbaryo,Gbaryo1,Pizza,Arbuzova}. Baryogenesis in the context of other theories such as in Lorentz symmetry violating models has also been studied \cite{baryoinLV}. 

The baryon asymmetry problem is usually referred to the observational number $\eta_B$, which is the ratio of the baryon number density, $n_B$, to the entropy density, $s$ (or the photon number density, $n_{\gamma}$). According to the latest measurement by the Planck satellite \cite{Planck2015}, the dimensionless baryon density $\Omega_B h^2= 0.02225\pm 0.00016$ gives the baryon asymmetry equal to
\be
\eta_B= \frac{n_B}{s} = (8.678\pm 0.062)\cdot10^{-11} ,
\ee
where $\eta_B= 3.9 \cdot 10^{-9}\Omega_B h^2$. The entropy density and the photon number density are related by $s\approx7.04 n_\gamma$.

In this paper we concentrate on the spontaneous baryogenesis approach with the baryon asymmetry generating fields being motivated by the dynamical physical constants. The paper is organised as follows. In Section \ref{spontan} we briefly sketch the idea of spontaneous baryogenesis. In Section \ref{dynbar}, which is the main body of this work, we discuss how to generate baryon asymmetry in spontaneous baryogenesis scenario where the role of a thermion is played by dynamical constants such as the varying gravitational constant $G$, varying fine structure constant $\alpha$, and the varying speed of light $c$. In Section \ref{discuss} we summarise our results and give conclusions.

\section{Spontaneously generated baryon asymmetry in the universe}
\label{spontan}

As it was mentioned in the Introduction, unlike the Sakharov baryogenesis, the spontaneous baryogenesis is based on two assumptions \cite{cohenkaplan}: 1) baryon number violating interactions appear in thermal equilibrium; 2) CPT is not an exact symmetry of the early universe since its expansion violates Lorentz symmetry and a time-reversal. There is a relation between the Hubble parameter evolution and the size of CPT violation which can be tighten to the ''effective'' baryon number violating interactions. These being initially large, after the universe cools down and can be approximated by zero temperature, become gradually negligible so that baryon number violating interactions become CPT invariant and then allow the Lorentz invariant vacuum as we observe now. 

The key point is to consider a scalar field $\varphi$---in the original approach called thermion \cite{cohenkaplan}---which spontaneously breaks the baryon symmetry by a term in the action
\be
{\cal L} = \lambda^2 (\partial_{\mu} \varphi) J^{\mu}_B \ ,
\label{BV1}
\ee
where $\lambda$ is a characteristic cut-off length scale of the spontaneous baryogenesis model ($l_{pl} \leqslant \lambda < l_{GUT}$; $l_{pl}$ is the Planck length and $l_{GUT}$ is the Grand Unified Theory length scale), $J^{\mu}_B$ is the baryon current, and the Greek indices run from 0 to 3. After integrating (\ref{BV1}) by parts,  one obtains 
\be
{\cal L} = \lambda^2 \varphi (\partial_{\mu} J^{\mu}_B)\ ,
\label{BV2}
\ee
which means that the baryon current $J^{\mu}_B$ cannot be conserved (or otherwise, $\partial_{\mu} J^{\mu}_B$ cannot be zero). 
If it was conserved, the baryon number would be preserved and so there was no baryon asymmetry in the universe. 
The underlying idea here is to replace the term $\partial_{\mu} J^{\mu}_B$ by some operator which violates the baryon number and additionally can also give rise to a decay of thermion field at late time to finally reach baryon conservation at the late universe. 

Considering a homogeneous and isotropic Friedmann Universe
\begin{equation}
    ds^2= -c^2dt^2 +a^2(t) \left[\frac{dr^2}{1-kr^2}+r^2 d\theta^2 +r^2 \sin^2\theta d\varphi^2 \right],  
    \label{FRW}
\end{equation}
($x^\nu= (x^0, x^1, x^2, x^3) = (ct, r, \theta, \varphi)$, $d/dx^0 = d/(cdt)$, $k=0, \pm 1$)
we can write down (\ref{BV1}) as 
\be
{\cal L} = \lambda^2 (\partial_0 \varphi) J^0_B \equiv \mu_B \Delta n_B 
\label{BV3}
\ee
where $\Delta n_B$ describes  the difference in the number density of particles and antiparticles: 
\begin{equation}
    J_B^0 = \Delta n_B =n_B - n_{\bar{B}} \ .
    \label{Stat_Mech_current_T}
\end{equation}
In fact, the term (\ref{BV3}) describes a CPT violating interaction which leads to different spectra for baryons and antibaryons. 
More precisely, the term (\ref{BV3}) breaks first, the CP symmetry and then the time symmetry due to having a nonzero vev $<\dot{\phi}>$ $\neq 0$, which finally leads to a CPT violation \cite{Gbaryo1}. 
The CPT symmetry ensures that particles and antiparticles equilibrate with the same thermal distribution, which is not a case when the symmetry is broken. 
 Therefore, the interaction (\ref{BV1}) or (\ref{BV2}) contributes to the Einstein equation by the energy--momentum tensor made out of these baryon number violating terms and shifts the energy of baryons with respect to the energy of antibaryons of about $2\mu_B$.  This shift is then interpreted as a chemical potential, which enters the particle/antiparticle Hamiltonian through the term \cite{DeSimone}
\begin{equation}
    \mu_B =E_B - E_{\bar{B}} \sim \lambda^2 \dot{\phi} ,  \label{pot_chem}
\end{equation} 
where $E_B$ is the energy of a baryon and $E_{\bar{B}}$ of an antibaryon. 
 For the antiparticles the chemical potential is $\mu_{\bar{B}}= - \mu_B $. 
 
 The thermodynamical quantities of some species ``$i$"---the number density $n_i$, the energy density, $\varepsilon_i = -\rho_i c^2$, and the pressure density $P_i$---are specified by the integrals over  their distribution functions:
 \begin{align}
    n_i&= \frac{g_i}{(2\pi \hbar)^3} \int f(\Vec{p})d^3\Vec{p}\ ,\\ 
    \varepsilon_i&= \frac{g_i}{(2\pi \hbar)^3} \int f(\Vec{p})E(\Vec{|p|})d^3\Vec{p}\ , \label{Stat_Mech_energy_density_integral}\\
    P_i&=\frac{g_i}{(2\pi \hbar)^3}  c \int f(\Vec{p}) \frac{|\Vec{p}|^2}{3E(|\Vec{p}|)} d^3\Vec{p}\ ,
\end{align}
 where $E$ is the energy, $\vec{p}$ is the momentum, $g_i$ is a number of the internal degrees of freedom, i.e. $g_i=2$ for a~photon, and $1/(2\pi \hbar)^3$ is a unit size of the phase space. The distribution function reads as \cite{DeSimone,perkins}:
\be 
    f(\Vec{p})= \frac{1}{(e^{(E_B-\mu_B)/k_B T} \pm 1)}\ 
\ee
 and due to the homogeneity and isotropy of the Friedmann universe it does not depend on the spatial coordinates and the momentum direction, so $f(\Vec{x}, \Vec{p}) \to f(p)$. 
 The sign ``$+$'' stands here for fermions (Fermi--Dirac statistics), ``$-$'' for bosons (Bose--Einstein statistics), and $k_B$ is the Boltzmann constant.  Given this, an excess of a baryon number over an antibaryon number (\ref{Stat_Mech_current_T}) can be written as
\begin{align}
    \Delta n_B& = \frac{g_i}{(2\pi^2 \hbar)^3}\times  \nonumber \\  
    &\int_0^\infty d^3\Vec{p} \left[\frac{1}{e^{(E_B-\mu_B)/k_B T} \pm 1 } 
    -  \frac{1}{e^{(E_B+\mu_B)/k_B T}\pm 1  }    \right] ,
\end{align}
which by using (\ref{Stat_Mech_current_T}) and (\ref{pot_chem}) gives an approximate result for the particle--antiparticle excess as 
 \begin{equation}
    J_B^0 =   \Delta n_B \simeq \frac{g_i}{6} \frac{k_B^2}{(\hbar c)^3}\mu_B T^2 \ .
    \label{Stat_Mech_baryon_excess_muT2}
    \end{equation}
The entropy density, $s$, for bosons (here: the radiation)  is given by:
    \begin{equation}
    s= g_{*s} \frac{2\pi^2}{45}\frac{k_B^4}{(\hbar c)^3 }T^3 \label{Stat_Mech_entropy_density}
    \end{equation}
 and $g_{*s}$ is the effective number of degrees of freedom, which differs from $g_*$ present in the solution of the integral (\ref{Stat_Mech_energy_density_integral}) for the energy density:
\begin{equation}
    \varepsilon_i =  g_{*} \frac{\pi^2}{30}\frac{k_B^4}{(\hbar c)^3 }T^4\ . \label{Stat_Mech_en_density_temp}
\end{equation}
  When all the species have the same temperature and the equation of state may be approximated by $p\simeq 1/3 \rho_i c^2$, these quantities appear equal, $g_{*s}=g_*$. 
 Since above the temperature $T\sim 200$ GeV all the particles are relativistic, we can find the value of $g_*=106.75$ by summing up their internal degrees of freedom \cite{DeSimone,perkins}.
 
 By combining equations (\ref{pot_chem}) and (\ref{Stat_Mech_baryon_excess_muT2}) we can write the final expression for the baryon asymmetry parameter:
\begin{align}
    \eta_B = \frac{\Delta n_B}{s}= \frac{15 g}{4\pi^2g_{*s}}  \frac{1}{k_B^2 T} \mu_B \ ,\label{eta_b}
\end{align}
 which has a dimension of (K/J) in SI units. 
 Another parameter describing the preference of matter over antimatter is the baryon to photon number ratio, $\eta_{B\gamma}= \Delta n_B / n_\gamma$. 
 However, until the photon decoupling (T $\sim 0.3$ eV) the photon density number $n_{\gamma}$ vary significantly throughout the epochs of the evolution of the Universe.
 For this reason, the entropy density $s$, which remains more or less constant at all energies, seems to be a better quantifier of the baryon asymmetry.

\section{Dynamical constants driven baryogenesis}
\label{dynbar}

An idea of varying physical constants is in a way analogous to the idea of running  coupling constants in quantum field theory, i.e. that there is some interaction due to perhaps unknown physics, which causes these constants to vary in time and possibly in space. 
In practice, what one does is that one replaces the constants of nature by some physical fields, which have their own dynamics. 
The first fully quantitative framework for this was developed for varying gravitational constant (as a coupling constant of gravitational interaction) by Jordan \cite{Jordan} and Brans--Dicke \cite{BD}. 
They were motivated by the earlier Large Number Hypothesis of Dirac \cite{Dirac1937}  being the consequence of even earlier ideas of Weyl \cite{Weyl} and Eddington \cite{Eddington}. 
Among the rich set of fundamental constants (for a review see Refs. \cite{uzan,UzanLR,Carlos17,barrowbook}) the series of them are subject to dynamical studies. These are the gravitational constant $G$ \cite{BD}, the proton to electron mass ratio $\mu = m_p/m_e$ \cite{mu}, the fine structure constant $\alpha = e^2/\hbar c$ \cite{alphaobs} ($\hbar$ is the Planck constant and $c$ is the speed of light) and related to this charge of an electron $e$ \cite{Bekenstein} or permittivity of vacuum $\epsilon_0$ \cite{BM2015}, and the velocity of light $c$ \cite{Moffat93A}. 

Though one usually considers the dynamics of the constants separately, the models in which two of the constants vary instantaneuosly have also been considered. Out of them the most natural are modified varying both $G$ and $c$ models \cite{Barrow99,varGc,adam2015} since these constants show up together in the Einstein-Hilbert action for gravity and in the Einstein field equations. In fact, they can be classified as an extension of Brans-Dicke models into a varying $c$ case. Another extension of this type which is based on Brans-Dicke model are varying both $G$ and $\alpha$ models \cite{varGalpha}. On the other hand, varying both $\alpha$ and $c$ models would not perhaps make so reasonable because $\alpha$ and $c$ are related via the definition of the fine structure constant and the effects of changes of these constants would have to be indicated separately in the construction. 

Our main idea here is to have the scalar fields which are responsible for the dynamics of the constants such as $G$, $\alpha$, and $c$ to play the role of a thermion in the spontaneous baryogenesis scenario as described in Section \ref{spontan}. 
In the following we will discuss all these three varying constants scenarios in that context.

\subsection{Dynamical gravitational constant $G$ driven baryogenesis} 
\label{G_chapter}

 The action for (varying $G$) Brans--Dicke theory reads as \cite{BD}:
\begin{equation}
    S= S_{BD} + S_m + S_B \ , \label{eq:lag_BD_baryo}
\end{equation}
 where:
\begin{align}
    S_{BD}&= \frac{c^3}{16 \pi } \int d^4x \sqrt{-g} \left( \phi R - \frac{\omega}{\phi} \partial_\mu \phi \partial^\mu \phi \right)\ , 
    \label{SBD} \\
    S_{m}&=\int d^4x \sqrt{-g}\mathcal{L}_{m}  \ , \label{Sm}\\
    S_B&=  \frac{c^3}{16 \pi }\int d^4x  \lambda_G^2 \partial_\mu \phi J_B^\mu\ , \label{eq:BD_baryo_term}
\end{align}
 where $S_{BD}$ is the standard Brans--Dicke action, $S_m$ is the matter action, $\phi$ is the Brans--Dicke scalar field, $\omega$ is a constant Brans--Dicke parameter. We have added an extra term, $S_B$,  which describes the interaction responsible for baryogenesis \cite{DeSimone}. The quantity $\lambda_G$ is the characteristic cut-off scale for $G$-varying baryogenesis models. 
 It is worth to mention that we have used the notation for the action (\ref{SBD}) with the speed of light being $c^3$ rather than $c^4$ in front of the integral. 
 Here we follow the notation of Refs.  \cite{LL,Will} compensating one $c$ to be kept in the definition of the null coordinate $x^0 = ct$ rather than $x^0 = t$ as in most of the textbooks (e.g. \cite{HE}).
 We will come back to this problem in section \ref{c_chapter} where the models with varying speed of light are considered (cf. also the detailed discussion of Ref. \cite{JCAP15}). 
  
The scalar field $\phi$ is related to the varying gravitational constant $G$ as
\begin{equation}
    \phi(x^\nu)= \frac{1}{G(x^\nu)}\ . \label{BD_field}
\end{equation}
 The action (\ref{eq:lag_BD_baryo}) varied with respect to the metric yields the field equations:
\begin{equation}
    G_\mu^{~\nu} = \frac{8 \pi\phi^{-1}}{c^4 } \left({T_m}_\mu^{~\nu}+ {T_{BD}}_\mu^{~\nu} + {T_{B}}_\mu^{~\nu} \right)\ , \nonumber
\end{equation}
 where the tensors ${T_i}_\mu^{~\nu}$ are given by 
\begin{eqnarray}
\label{mattertrafo}
    {T_{m}}_{\mu}^{~\nu} &=& g_{\mu\sigma}\frac{2}{\sqrt{-g}}\frac{\partial}{\partial  g_{\sigma\nu}}\left(
    \sqrt{-g}\mathcal{L}_{ m}\right)~,\\
  \label{TBD}
    {T_{BD}}_\mu^{~\nu}&=&\frac{c^4}{8 \pi}\left(\nabla_{\mu}\nabla^{\nu} \phi - \delta_{\mu}^{\nu}\Box \phi \right) \\
    &+& \frac{c^4}{8 \pi}\left( \partial_{\mu} \phi \partial^{\nu} \phi - \frac{1}{2} \delta_{\mu}^{\nu} \partial_{\beta} \phi \partial^{\beta} \phi \right) ,\nonumber \\
    {T_{B}}_\mu^{~\nu}&=&   \frac{c^4}{16 \pi }\lambda_G^2 \delta_\mu^\nu  \partial_\sigma \phi J_B ^\sigma\ ,\label{BD_baryo_en_density}
\end{eqnarray}
 and the equation of motion of the field $\phi$ takes the form:
\begin{align}
    \Box \phi &= \frac{8 \pi}{c^4 (3+2\omega)} T_m  \label{BD_baryo_eq.of.motion}\\ 
   & +\frac{\lambda_G^2}{3+2 \omega} \phi \left( \partial_\mu J_B^\mu + J_B^\gamma {\Gamma^\mu}_{\mu \gamma } + \frac{2}{\phi} \partial_\mu \phi J_B^\mu     \right)\nonumber\ ,
\end{align}
where ${\Gamma^\mu}_{\mu \gamma }$ are the Christoffel connection coefficients. 
 Assuming that the field (\ref{BD_field}) is homogeneous and isotropic we can write the Friedmann equation for the flat Universe as follows:
\begin{align}
    H^2 = \frac{8 \pi }{3\phi}\rho_m  -H H_\phi+ \frac{\omega}{6}H_\phi^2 +\frac{c\lambda_G^2}{6}H_\phi J_B^0 ,\label{BD_baryo_friedman_eq}
\end{align}
where $H$ denotes the Hubble parameter, $H_\phi= \dot{\phi}/\phi$ is the rate of variation of $G(t)$ and  $\rho_m$ is the matter energy density with corresponding pressure, $p_m$. The acceleration equation is given by:
\begin{align}
     \frac{\ddot{a}}{a}&= -\frac{4 \pi}{3\phi}
    \left(\rho_m+\frac{3p_m}{c^2}\right)-\frac{1}{2}HH_{\phi}
    - \frac{1}{6}(3+2\omega)H_{\phi}^2
    -\frac{1}{2}\dot{H}_{\phi}\nonumber\\
    &- \frac{c\lambda_G^2}{3}H_\phi J_B^0 \ ,
   \label{BD_acceleration_eq}
\end{align}
 In order to calculate (\ref{eta_b}), we need to solve the equation of motion (\ref{BD_baryo_eq.of.motion}), which for Friedmann metric takes the form:
\begin{align}
    \ddot{\phi}+3H\dot{\phi} &= \frac{8 \pi}{ (3+2\omega)} (1-3 w)\rho_m \nonumber\\ 
    & -\frac{\lambda_G^2 c^2}{3+2 \omega} \phi \left[\dot{J}_B^0 +3H J_B^0 +2\frac{\dot{\phi}}{\phi} J_B^0     \right]\ \label{BD_baryo_eq.of.motion2},
\end{align}
 where $w$ is an index of the barotropic equation of state $p= w \rho_m c^2$, and $p$ is the pressure. 
 
The main problem with the set of equations (\ref{BD_baryo_friedman_eq})-(\ref{BD_baryo_eq.of.motion2}) (compare Ref. \cite{BDsolns} for example) is that in radiation dominated universe the first term on the right-hand side of equation (\ref{BD_baryo_eq.of.motion2}) vanishes. This, after additionally neglecting the last term of baryogenesis which is in fact small, leads to a pure scalar field (or stiff-fluid) domination with a simple integral 
 \be
 \dot{\phi} \propto a^{-3}(t) .
 \label{SF}
 \ee
 Despite this solves easily, still the solutions of the whole set of equations for $a(t)$ and $\phi(t)$ are non-trivial. In fact, one can postulate the power-law solutions which would include both an early universe "scalar field domination" and late universe radiation domination together. However, as it has been shown in Ref. \cite{BDdust} that the unique power-law solutions which allow current acceleration of the universe $a \propto t^{4/3}$, $\phi \propto t^{-2}$  are possible for the dust $(p=0)$ models if one also adds a specific scalar field potential. Because the matter applied is dust, then the relation (\ref{SF}) is modified accordingly.  
 
In fact, if one neglected radiation also in the Friedmann equation (\ref{BD_baryo_friedman_eq}), then one would get a unique solution which would pick up some specific powers in the scale factor and scalar field power-law time  dependence. However, in our case we deal with the early universe and so radiation is the crucial component. Because of that we need to rely on the relation (\ref{SF}), though modified slightly by the baryogenesis term which is pretty small. The solutions which in fact keep relation (\ref{SF}) valid but also include radiation have been studied in Ref. \cite{Morganstern}. In particular, it was found that there exist two regimes in which simple power-law solutions (with radiation and the scalar field present) exist. One of them applies close to a big-bang $a \to 0$, where the scalar field is dominating (behaving as a stiff-fluid), and another to the late time evolution $a \to \infty$, when the radiation comes to dominate. The first solution gives a simple power law for the scale factor $a \propto t^{1/3}$, while the second gives standard radiation-dominated power law behaviour $a \propto t^{1/2}$. 

What is crucial here is that there exist solutions fulfilling set of equations (\ref{BD_baryo_friedman_eq})-(\ref{BD_baryo_eq.of.motion2}) which include radiation and asymptote from the power-law solution $a \propto t^{1/3}$ to possibly another power law solution $a \propto t^{1/2}$. They can be parametrised by the values of some extra parameter  which takes some specific value for the asymptotic $a \propto t^{1/3}$ solution. In other words, we can consider the solutions which are in stiff-fluid regime, but which are slightly modified by the presence of radiation. Such an approach have been applied to Brans-Dicke theory already in Refs. \cite{ChenBDBaryo,Li} though only some simple examples of baryogenesis out of the whole set of admissible values of the extra parameter (in Ref. \cite{ChenBDBaryo} parameter $n$) have been studied. 
 
In the following we will explain the above approach step by step.
 
 We start with the value of the chemical potential (\ref{pot_chem}) which now reads as:
 \begin{equation}
     \mu_B= \frac{c^3}{16\pi}\lambda_G^2 \dot{\phi} \ . \label{BD_baryo_pot_chem}
 \end{equation}
 Given this, we find (\ref{Stat_Mech_baryon_excess_muT2}) for the varying $G$ case, which is:
 \begin{equation}
     J_B^0=\frac{g_i\lambda_G^2}{96\pi \hbar^3}k_B^2\dot{\phi}T^2 \label{BD_J0_nB} \ , 
 \end{equation}
 and in the next step, by inserting (\ref{BD_J0_nB}) into (\ref{BD_baryo_eq.of.motion2}) we obtain the modified with baryogenesis term equation of motion for the scalar field:
\begin{equation}
    \ddot{\phi}+3H\dot{\phi}= -\frac{2\beta (T^2{\dot{\phi}}^2 +  T \dot{T} \phi \dot{\phi})}{1+\beta T^2 \phi}\ ,\label{BD_baryo_eom3}
\end{equation}
 where $\beta$ is constant with the dimension of $m^5 K^{-2} s^{-4} J^{-1}$:
 \begin{equation}
    \beta= \frac{c^4}{16 \pi(3+2\omega) } \frac{g_i}{6} \frac{k_B^2}{(\hbar c)^3} \lambda_G^4 \ . \label{BD_beta}
\end{equation}
 The equation (\ref{BD_baryo_eom3}) can be solved, but as it can be proven numerically, the contribution from the right-hand side is small because during baryogenesis the temperature does not change significantly, i.e. $\dot{T} \approx 0$. Besides, for the length scale $\lambda_G$  in the range $l_{Pl}\sim 10^{-35}$ m $<\lambda_G < l_{GUT}\sim 10^{-31} $m, $\beta T^2\ll 1$ for considered temperatures  $10^{13}$ GeV $< T < 10^{15}$ GeV, and for $500<\omega<40 000$. Consequently, the right hand side of (\ref{BD_baryo_eom3}) can be neglected and the equation of motion simplifies to
 \begin{equation}
     \ddot{\phi}+ 3H \dot{\phi}=0  \label{BD_eom_fully_simplified} \ ,
 \end{equation}
 which immediately gives the solution (\ref{SF}). Without any loss of generality in looking for the power law solutions we will describe the dynamics of the field $\phi$ by making an ansatz for the variability of the gravitational constant $G$: 
 \begin{equation}
     \frac{1}{\phi(t)} = G(t)= G(t_0)\left[\frac{a(t)}{a(t_0)}\right]^q \ ,\label{BD_G_ansatz}
 \end{equation}
 where a dimensionless parameter q is a measure of the variation of $G(t) \equiv G$, and $a(t) \equiv a$, $a(t_0) \equiv a_0$ are the scale factors at times $t$ and $t_0$ . For $q=0$, $G(t)$ is equal to a currently measured value of the gravitational constant, $G(t_0)= G_0$. The parameter $q$ (which is equivalent to a parameter $n=1/(3-q)$ of Ref. \cite{ChenBDBaryo}) also measures the deviation from $a \propto t^{1/3}$ as mentioned earlier.  Applying the ansatz (\ref{BD_G_ansatz}) into (\ref{BD_eom_fully_simplified}) yields 
 \begin{equation}
     \frac{\ddot{a}}{\dot{a}}+ (2-q)\frac{\dot{a}}{a}= 0\ , 
 \end{equation}
 which has the following solution:
\begin{align}
      a(t)&= a_{in} \left[ 1+(3-q)(t-t_{in})H_{in} \right]^{\frac{1}{3-q}} \ , \label{BD_a}\\
     \dot{a}(t)&= \dot{a}_{in} \left[ 1+(3-q)(t-t_{in})H_{in} \right]^{\frac{q-2}{3-q}}\ \label{BD_a_dot}
\end{align}
 and allows us to find the Hubble parameter $H$:
 \begin{align}
     H(t)= H_{in}  \left[ 1+(3-q)(t-t_{in})H_{in} \right]^{-1}\ , \label{BD_Hubble_par}
 \end{align}
 which then gives the values of $\phi$ and its derivative $\dot{\phi}$ as the functions of time:
\begin{align}
    \phi(t)&= \phi_{in}  \left[ 1+(3-q)(t-t_{in})H_{in} \right]^{-\frac{q}{3-q}}\ , \label{BD_phi_time}\\
    \dot{\phi}(t)&= -q \phi_{in} H_{in} \left[ 1+(3-q)(t-t_{in})H_{in} \right]^{-\frac{3}{3-q}}\ . \label{BD_phi_dot_time} 
\end{align}
The indices ``$in$" denote the initial values of the scale factor $a_{in}$, the Hubble parameter $H_{in}=\dot{a}_{in}/a_{in}$ and the field $\phi_{in} = 1/G_{in}$ at the beginning of baryogenesis, when $t = t_{in}$. It is worth mentioning that in the limit $q \to 0$ the above solutions give an asymptotic early time behaviour $a \propto t^{1/3}$ as it should be following the work of Ref. \cite{Morganstern}. 

In fact, the scale factor (\ref{BD_a}) is a superposition of a solution for pure radiation and for pure stiff fluid which mimics the scalar field. In the above mentioned limit $q\to 0$, the solutions (\ref{BD_a}) and (\ref{BD_phi_time})
 solve simultaneously the whole set of the field equations (\ref{BD_baryo_friedman_eq}), (\ref{BD_acceleration_eq}), (\ref{BD_eom_fully_simplified}), when satisfied:
 \begin{equation}
     \rho_{m0}= \frac{3\phi_{in}}{8 \pi} \left(\frac{a_{in}}{a_0} \right)^{6} \left( 2 H_{in} \frac{3-q}{6-q}\right)^2 \left(1-q-\frac{\omega}{6}q^2 \right)  \ .\label{BD_en_den_constant1}
 \end{equation} for the radiation energy density taking form in the Brans--Dicke theory:
\begin{align}
    \rho_m= \rho_{m0}\left( \frac{a_0}{a}\right)^6 \ .
\end{align}
Notice that in order to get (\ref{BD_eom_fully_simplified}) the contribution (\ref{eq:BD_baryo_term}) from baryogenesis term has been neglected in the Friedmann equation for the same reason for which we neglected it in the equation of motion (\ref{BD_eom_fully_simplified}). In fact, if we have also dropped radiation contribution in the Friedmann equation so $\rho_{m0} =0$, we would obtain the condition linking the values of $q$-parameter and $\omega$: $q=(\sqrt{3}/\omega)(-\sqrt{3} \pm \sqrt{2\omega + 3})$ which would restrict the freedom of choice of $q$ in a similar way as in the dust case considered in Ref. \cite{BDdust}.

 In order to define the temperature dependence of $\phi$ we combine equations (\ref{BD_baryo_friedman_eq}), (\ref{Stat_Mech_en_density_temp}), (\ref{BD_Hubble_par}), (\ref{BD_phi_time}), and (\ref{BD_phi_dot_time}). 
 Due to the fact that the right hand side of (\ref{BD_baryo_eq.of.motion2}) is negligible, we feel excused to neglect the term $c\lambda_G^2/6 H_\phi J_B^0$ in the Friedmann equation.
    \begin{figure}[t]
     \includegraphics[width=8.5cm]{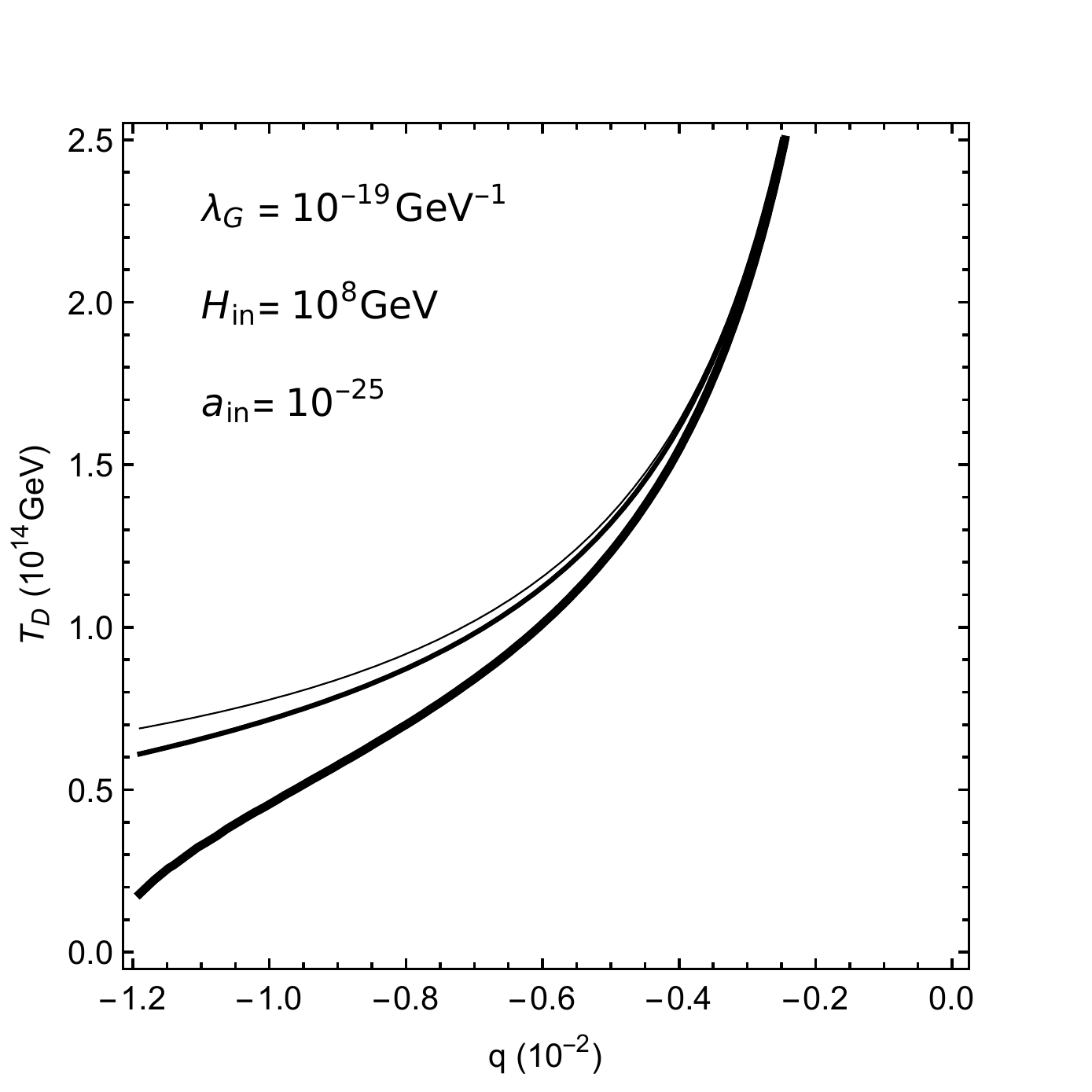}
     \caption{The decoupling  temperature, $T_D$, as function of the parameter $q$, in the model with varying $G$ using the currently measured value of the asymmetry, $\eta_B~\simeq~8.6~\cdot~10^{-11}$.  The thin line corresponds to  $\omega=1~000$, the middle line corresponds to  $\omega=10~000$, and the thick line to $\omega=40~000$.
   The plots were made for  $\lambda_G =10^{-19}$ Gev$^{-1}$, $a_{in} = 10^{-25}$, and $H_{in} = 10^{8}$~GeV.}  \label{fig:G-TD_q_for_omega}
\end{figure}
 This gives a time--temperature relation in our model as
\begin{align}
    &\left[1+(3-q)(t-t_{in})H_{in}\right]^{\frac{6-q}{3-q}}=\nonumber \\
    &\phi_{in} H_{in}^2 \left( 1-q-\frac{\omega}{6} q^2  \right) \left[ 
      \frac{8 \pi ^3 g_*}{90 c^2} \frac{k_B^4}{(\hbar c)^3} \right]^{-1}  T^{-4}\ , \label{BD_time_temp_relation}
\end{align}
 which can be implemented into (\ref{BD_phi_dot_time}) and together with eq. (\ref{eta_b}) and (\ref{BD_baryo_pot_chem}) gives the final expression for the baryon asymmetry: 
 \begin{align}
    \eta_B &=  -q\phi_{in} H_{in}\frac{15c^3}{16\pi } \frac{\lambda_G^2 g_i}{8\pi^2 g_{*}} \frac{1}{k_B^2 T} \quad \times\label{BD_eta_b_final}\\
    &\left[ \phi_{in} H_{in}^2 \left( 1-q-\frac{\omega}{6}q^2  \right) \left( 
      \frac{8 \pi ^3g_* }{90 c^2} \frac{k_B^4}{(\hbar c)^3} \right)^{-1}  T^{-4} \right]^{-\frac{3}{6-q}}\nonumber \ . 
\end{align}
\begin{figure}[t!]
\centering
    \includegraphics[width=8.5cm]{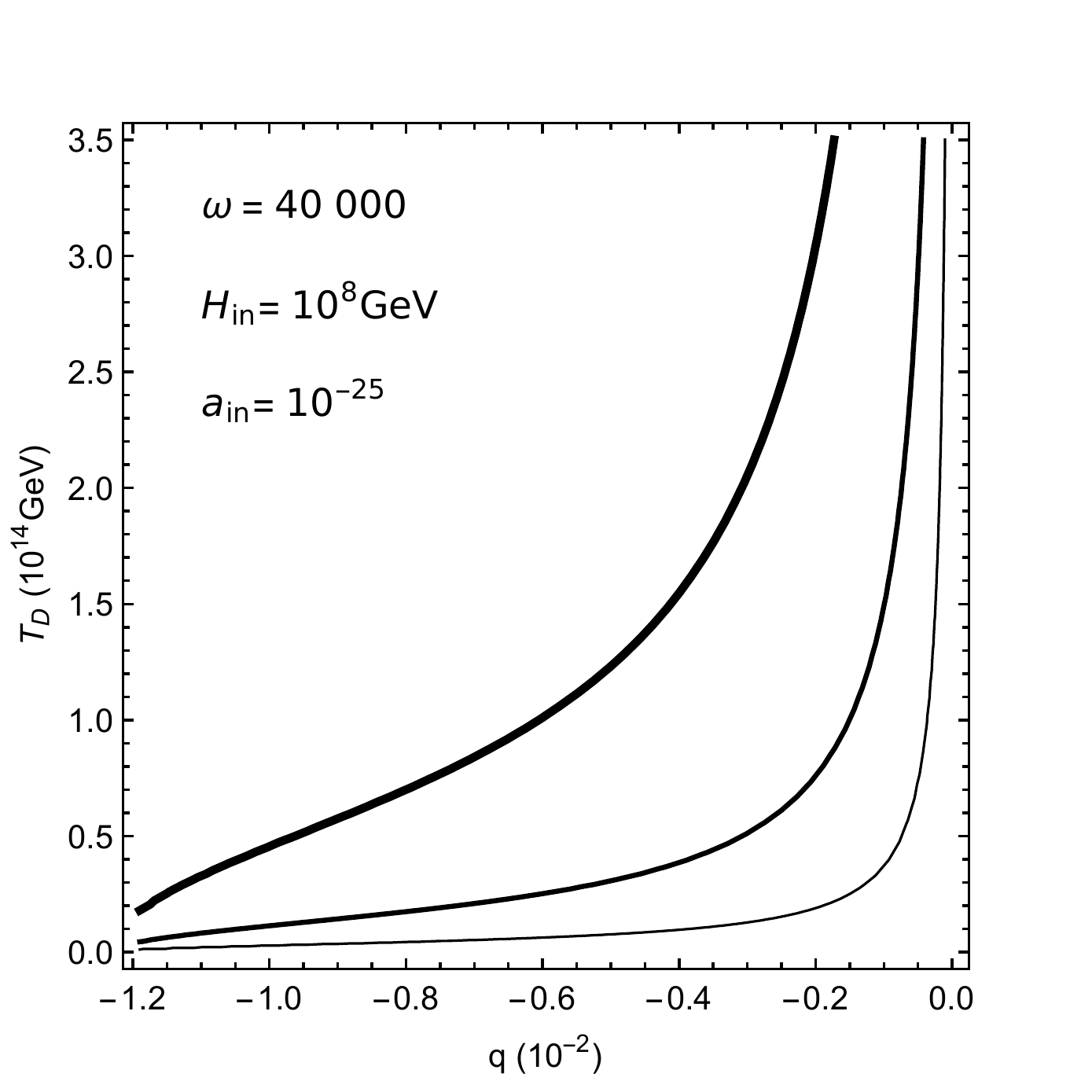}
  \caption{The decoupling  temperature, $T_D$, as function of the parameter $q$, in the model with varying $G$ using the currently measured value of the asymmetry, $\eta_B~\simeq~8.6~\cdot~10^{-11}$.
   The thick line corresponds to $\lambda_G= 10^{-19}$~GeV$^{-1}$, the middle line corresponds to $\lambda_G=2\cdot 10^{-19}$ GeV$^{-1}$, and the thin line to $\lambda_G= 4\cdot10^{-19}$ GeV$^{-1}$. The plots were made for $\omega= 40~000$, $a_{in} = 10^{-25}$, and $H_{in} = 10^{8}$ GeV.} \label{fig:G-TD_q_for_lambda}
   \end{figure}
 \begin{table}[b]
\centering
\caption{Limits for the parameter $q$ for some specific values of  $\omega$ ($\omega = -3/2$ for conformal relativity, $\omega = -1$ for superstring theory, $\omega = \infty$ for Einstein gravity limit). }
\label{tab:G-q_limits}
\begin{tabular}{c|c}
\hline\hline
 \rule{0pt}{15pt}$\omega$ & $q $ \\[7pt] \hline\hline
\rule{0pt}{10pt}
-1.5      &   2                                      \\[3pt]
-1         &  (-$\infty$; $3 - \sqrt{3}$) $\cup$ (3 + $\sqrt{3}$; $\infty$) \\[3pt]
1 000    & (-0.0805; 0.0745)    \\[3pt]
5 000    & (-0.0352; 0.0340)                     \\[3pt]
10 000   & (-0.0248; 0.0242)                     \\[3pt]
40 000   & (-0.0123; 0.0122)                    \\ [3pt]
$\infty$ & 0 \\\hline
\end{tabular}
\end{table}
The quadratic equation in  (\ref{BD_time_temp_relation}) relates $\omega$ with the parameter $q$ and gives a bound on the allowed values of the field $\phi$. For $\omega > 0$ the bound is given by
\begin{equation}
   q \in \left(\frac{-3-\sqrt{9+6\omega}}{\omega};    \frac{-3+\sqrt{9+6\omega}}{\omega}\right)\ ,
   \label{limitq}
\end{equation}
while for $\omega < 0$ it reads as 
 \begin{figure*}[ht!]
  \includegraphics[width=8.5cm]{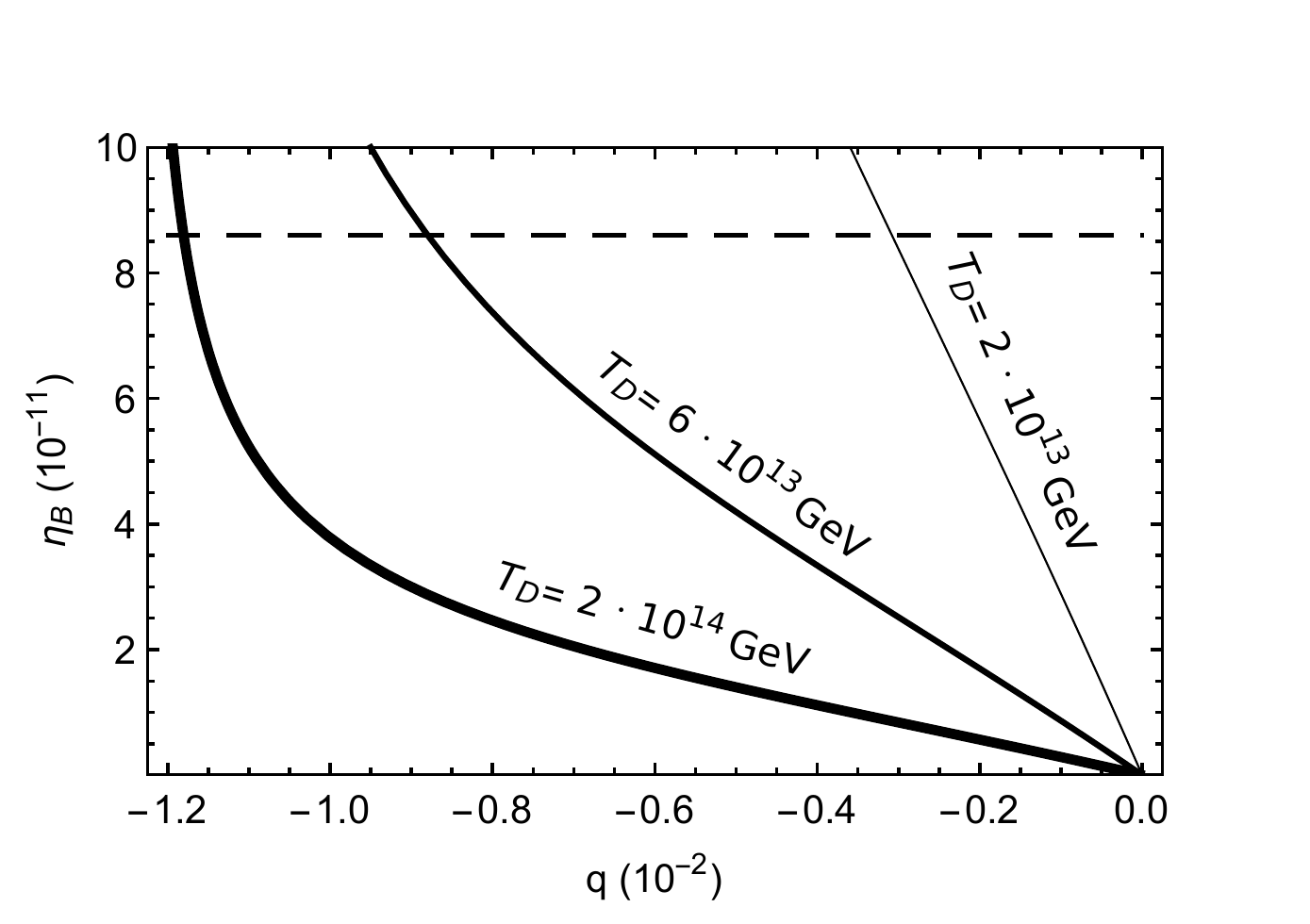}
    \includegraphics[width=8.5cm]{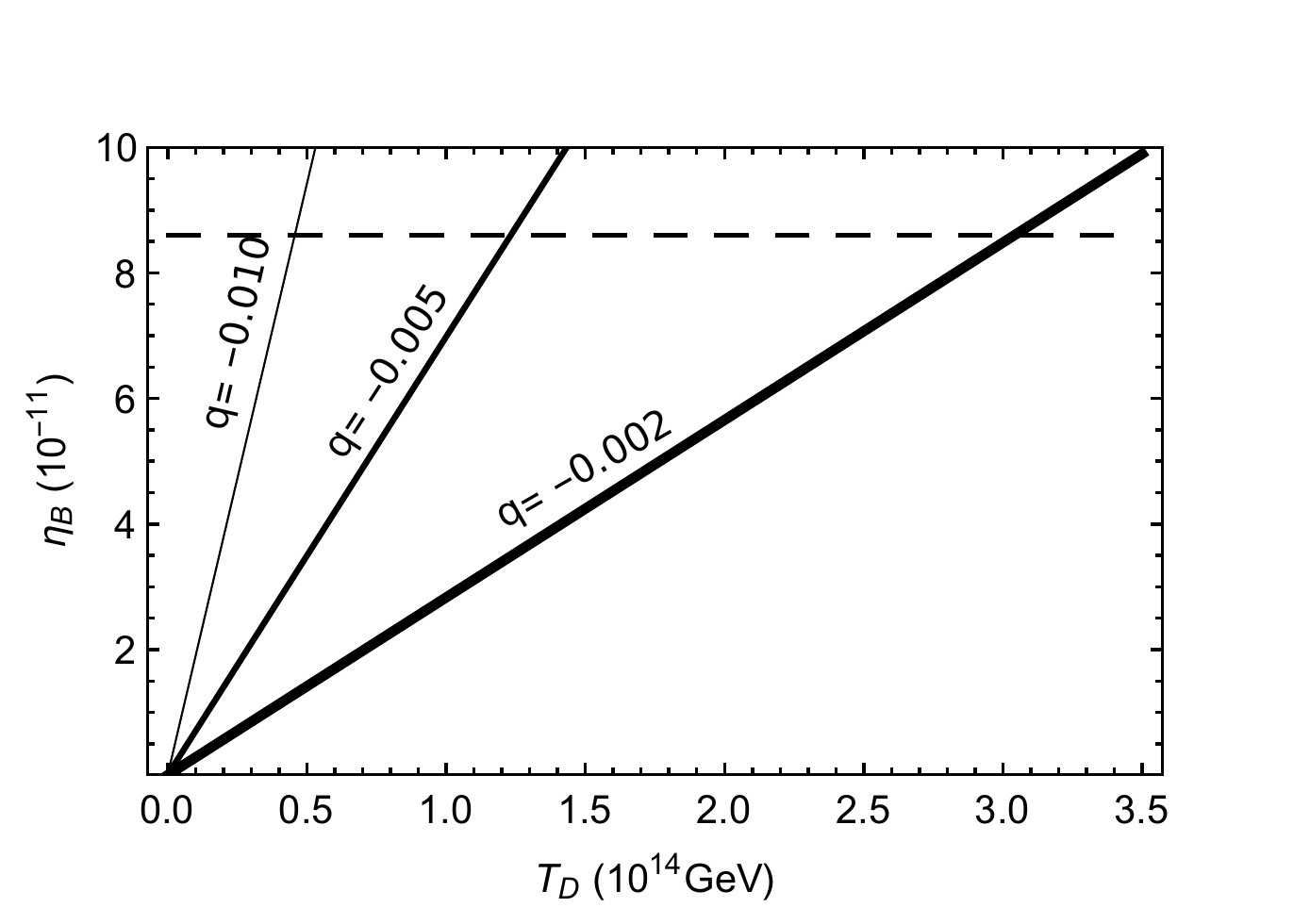}
 \caption{The baryon to entropy ratio $\eta_B$ (\ref{BD_eta_b_final}) as a function of the parameter $q$ (left) and the decoupling temperature $T_D$ (right) in the varying $G$ model. The horizontal dashed lines indicate the currently measured value of the asymmetry, $\eta_B\simeq 8.6\cdot10^{-11}$. 
  The plot on the left was made for three different values of $T_D$. The thick line corresponds to temperature $T_D= 2\cdot10^{13}$~GeV, the line in the middle to $T_D= 6\cdot10^{13}$~GeV and the thin one to $T_D=~2\cdot10^{14}$~GeV. 
  The plot on the right presents the relation $\eta_B(T_D)$ for three values of the parameter $q$. The thick line  corresponds to $q= -0.002$, the line in the middle to $q= -0.005$,  and the thin line to $q=-0.010$.
  Both plots were made with the initial conditions for baryogenesis: $a_{in}~=~10^{-25}$, $H_{in} = 10^{8}$ GeV, $\omega = 40~000$, $\lambda = 10^{-19}$ GeV$^{-1}$.
  }\label{fig:G-eta_q_TD}
\end{figure*}
\begin{equation}
   q \in \left(- \infty; \frac{-3+\sqrt{9+6\omega}}{\omega}\right) \cup \left(\frac{-3-\sqrt{9+6\omega}}{\omega}; +\infty \right).
   \label{limitq2}
\end{equation}
The limits for $q$ for some specific $\omega$ has been listed in TABLE \ref{tab:G-q_limits}. In order to be consistent with the current measurements, we take  $\omega=40~000$ and the corresponding limit for $q$ \cite{Bertotti,Will_Living,Ooba}. Despite the fact that $\omega$ could have been smaller in the early Universe, taking $\omega=40~000$ seems to be justified by the fact that it has small impact on the decoupling temperature $T_D$ in (\ref{BD_eta_b_final}).

We limit ourselves to consider the negative values of $q$, only ($G$ decreases during the evolution of the universe).
The positive values of $q$ would result in negative $\eta_B$, hence a universe with an excess of antimatter on matter. 
The currently measured value of $\omega$ imposes even stronger bound onto the value of $q$ and narrows the limit to the range $(0; 0.0122)$ (see TABLE \ref{tab:G-q_limits}). 
Nevertheless, the parameter $\omega$ does not have a strong influence onto $T_D$ for the values of $q$ taken from this range. 
Its influence becomes more pronounced for smaller $q$ (see Fig. \ref{fig:G-TD_q_for_omega}).
An interesting observation from (\ref{limitq2}) is that for the well-known from the literature case---the conformally invariant gravity \cite{adp07}, $\omega = -3/2$---an allowed value of $q$  is positive, so this theory seems to contradict observed baryon asymmetry. On the other hand, the limit of $q$ for the low-energy superstring gravity \cite{superstring}, $\omega=-1$, allows creation of the observed baryon asymmetry in a $G$-varying universe. 
Note, that in the general relativistic limit, $\omega \to \infty$, the parameter $q$ vanishes, and according to (\ref{BD_G_ansatz}), the value of $G$ remains constant as it should be. 

We have found that the Brans--Dicke baryogenesis gives a currently measured value of the baryon asymmetry, $\eta_B$, for large range of the parameters $q$ and $T_D$. In order to be consistent with the results of BBN $(-0.10< \Delta G/G<0.13)$ \cite{UzanLR}), we have taken $q \sim 10^{-3}$. 
This results in the decoupling temperature $T_D\sim10^{14}$GeV. 
The measurements of CMB indicate that $q$ should be rather of the order of $10^{-2}$ $(-0.083< \Delta G/G<0.095)$, which corresponds to a change of $G$ between the recombination ($z\approx10^3$), and today ($z=0$) \cite{UzanLR}.  
In fact the parameter $q$ characterising the dynamics of $G$ can be calculated from the formula: 
 \begin{equation}
    q= -\log_{1+z_G} \left(1+\frac{\Delta G}{G}\right),
 \end{equation}
 where $z_G$ is a corresponding value of the redshift for which $\Delta G/G$ was measured. 
 
 We have performed calculations and plotted the results with the initial condition for the scale factor $a_{in}= 10^{-25}$ and the corresponding Hubble parameter $H_{in}= 10^{8}$ (calculated from $\Lambda$CDM). However, shifting the beginning of baryogenesis even from $a_{in}= 10^{-25}$ to $a_{in}= 10^{-30}$ only slightly changes $\eta_B$. Therefore, the second most relevant parameter to drive baryogenesis is the fundamental length, $\lambda_G$. 
In this case, a small change in the value of $\lambda_G$ results in a big change in the decoupling temperature, $T_D$ (see Fig. \ref{fig:G-TD_q_for_lambda}). 
The baryon to entropy ratio $\eta_B$ (\ref{BD_eta_b_final}) as a function of the parameter $q$ for three possible values of the temperature $T_D$ (see the plot on the left), as well as a function of temperature $T_D$ for three values of $q$ (see the plot on the right) has been presented in Fig. \ref{fig:G-eta_q_TD}.

\subsection{Dynamical fine structure $\alpha$ driven baryogenesis}
\label{al_chapter}

In this section we examine the application of the mechanism of spontaneous baryogenesis to the Bekenstein--Sandvik--Barrow--Magueijo (BSBM) \cite{SBM} model of the varying fine structure constant $\alpha$. Such models were first proposed by Teller \cite{Teller1948}, and later by Gamow \cite{Gamow67}, following the original path of the Large Number Hypothesis by Dirac \cite{Dirac1937}. A fully quantitative framework was developed by Bekenstein \cite{Bekenstein} in which a change in the fine structure constant $\alpha$ was fully identified with a variation of the constant electric charge, $e_0$ (cf. also Ref. \cite{BarrowAdP}). By assuming that $\alpha$ can vary, we also assume that the electric charge become space--time dependent. This gives a path to a charge conservation, but maintains the Lorentz invariance, which is usually violated in the theories of varying $\alpha$, where $e$ and $\hbar$ are kept constant, and $c$ varies. The electric charge variability was introduced by defining a dimensionless scalar field, $\epsilon(x^\mu)$, and as a consequence, $e_0$ was replaced by $e=e_0\epsilon(x^\mu)$. The electromagnetic tensor was then redefined to the form 
\begin{equation}
    F_{\mu\nu}=[(\epsilon A_\nu)_{'\mu}-(\epsilon A_\mu)_{'\nu}]/\epsilon \ ,\nonumber
\end{equation}
where the standard form of it can be restored for the constant $\epsilon$. For simplicity, in \cite{SBM,BarrowAdP} an auxiliary gauge potential, $a_\mu=\epsilon A_\mu$, and the electromagnetic field strength tensor, $f_{\mu\nu}= \epsilon F_{\mu\nu}$, were introduced, as well as a variable change: $\epsilon \to \psi \equiv \ln\epsilon$ was performed. The field $\psi$ in this model couples only to the electromagnetic energy, disturbing neither the strong, nor the electroweak charges, nor the particle masses.  

 The BSBM baryogenesis action is composed of
 \begin{align}
    S= S_g + S_{\psi}+S_{em}+ S_B
 \end{align} 
 and 
 \begin{align}
    S_g&= \frac{c^3}{16\pi G}\int d^4x \sqrt{-g}R\ , \\
    S_{\psi}&=\frac{1}{c}\int d^4x \sqrt{-g} \left( -\frac{\Omega}{2}\partial_\mu \psi \partial^\mu \psi\right) \ , \\
    S_{em}&=\frac{1}{c}\int d^4x \sqrt{-g}\left(-\frac{1}{4} f_{\mu\nu}f^{\mu\nu} \right)e^{-2\psi}\ ,  \\
    S_B&= \frac{c^3}{16\pi G}\int d^4x \sqrt{-g} \lambda_\alpha^2 (\partial_\mu \psi) J_B^\mu\ ,\label{BSBM_action_S_B}
\end{align}
 where $S_g$is  the gravitational action, $S_{em}$ is the electromagnetic part of the theory with the kinetic term $S_{\psi}$ and $S_B$ is the baryogenesis term with the field $\psi $ derivatively coupled to the baryon current $J_B^\mu$. 
 Similar to the original BSBM theory, the coupling constant $\Omega=\hbar c / \lambda $, is a constant introduced for dimensional reason (J/m), where $\lambda$ is considered the length scale of the electromagnetic part of the theory.  The constant $\lambda_\alpha$ is a cut-off length scale of the spontaneous baryogenesis model and is taken to be $\lambda_{Pl}<\lambda_\alpha<\lambda_{GUT}$. The field $\psi$ is given by:
\begin{equation}
    \psi= \frac{1}{2}\ln\left| \frac{\alpha}{\alpha_0}\right| \label{BSBM_psi1}
\end{equation}
 and is dimensionless. The field equations read as
\begin{equation}
    G_\mu^{~\nu} = \frac{8 \pi G}{c^4 } \left({T_{em}}_\mu^{~\nu}+ {T_{\psi}}_\mu^{~\nu} + {T_{B}}_\mu^{~\nu} \right)\ , 
\end{equation}
 where the tensors ${T_i}_{\mu}^{~\nu}$ are given by: 
\begin{align}
    {T_{\psi}}_{\mu}^{~\nu}&=  \Omega  \partial_\nu \psi \partial^\nu \psi - \frac{1}{2}\Omega \delta_\mu^\nu \left(  \partial_\beta \psi \partial^\beta \psi \right)^2 \label{BSBM_T_ps} ,\\
     {T_{em}}_\mu^{~\nu}&= -\left(\frac{1}{4} \delta_\mu^\nu f_{\alpha\beta} f^{\alpha\beta} - g^{\sigma\nu} f_{\sigma\beta}f_{\mu}^\beta \right)e^{-2\psi} \label{BSBM_T_em}\ ,\\
    {T_{B}}_\mu^{~\nu}&= \frac{c^4}{16 \pi G }\lambda_\alpha^2 \delta_\mu^\nu  \partial_\sigma \psi J_B ^\sigma\ .\label{BSBM_T_B}
\end{align}
and the equation of motion of the field $\psi$ is: 
\begin{equation}
    \Box \psi = \frac{2}{\Omega} e^{2\psi} \mathcal{ L}_{em} +\frac{c^4}{16 \pi G} \frac{\lambda_\alpha^2}{\Omega} \left( \partial_\mu J_B^\mu + J_B^\gamma \Gamma^\mu _{\mu \gamma}  \right)\ . \label{BSBM_eom_1}
\end{equation}

 The Friedmann equation for the flat Friedmann metric (\ref{FRW}) and the homogeneous field ansatz $\psi = \psi(t)$ reads as
\begin{align}
    H^2= \frac{8\pi G}{3}\left(  \rho_{em} + \frac{\Omega}{2c^4} \dot{\psi}^2\right) + \frac{c\lambda^2_\alpha}{6} \dot{\psi}J^0_B\  \ ,\label{BSBM_Friedmann_eq}
\end{align}
 where  $\rho_{em}$  is the electromagnetic field energy density which will be, later on, re-scaled as follows:  
 \begin{align}
    {T_{em}}_0^{~0} &= - \rho_{em} c^2= -\tilde{\rho}_{em}c^2e^{-2\psi}\ ,\label{BSBM_T_em0}
\end{align}
The acceleration equation is given by:
\begin{align}
    \frac{\ddot{a}}{a}= & -\frac{4 \pi G}{3}\left(\rho_{em}+\frac{3p_{em}}{c^2} \right) - \frac{8 \pi G}{3 c^4}\Omega\dot{\psi}^2- \frac{c\lambda_\alpha^2}{3}\dot{\psi}J^0_B\ ,
\end{align}
where  $p_{em}$ is  the electromagnetic pressure which  we re-scale  as:
\begin{align*}
{T_{em}}_1^{~1}= p_{em}=\tilde{p}_{em}e^{-2\psi} \ .
\end{align*}
 Similarly to the section \ref{G_chapter}, where the varying $G$ baryogenesis was discussed, the interaction (\ref{BSBM_action_S_B}) violates the CPT symmetry.  This results in a different thermal distributions for particles and antiparticles and contributes to the stress--energy tensor. this contribution may be  
 understood as a chemical potential:
\begin{align}
    \mu = \frac{c^3}{16\pi G} \lambda_\alpha^2 \dot{\psi}\ .\label{BSBM_pot_chem} \quad 
\end{align}
This together with (\ref{eta_b}) leads to the baryon to entropy ratio in the form:
\begin{align}
    \eta_b = \frac{\Delta n_B}{s}=  \frac{15c^3}{16\pi G} \frac{\lambda^2_\alpha g_i}{8\pi^2 g_{*s}} \frac{1}{k_B^2 T} \dot{\psi} \ .\quad  \label{BSBM_eta_B}
\end{align} 
\begin{figure}
   \includegraphics[width=8.5cm]{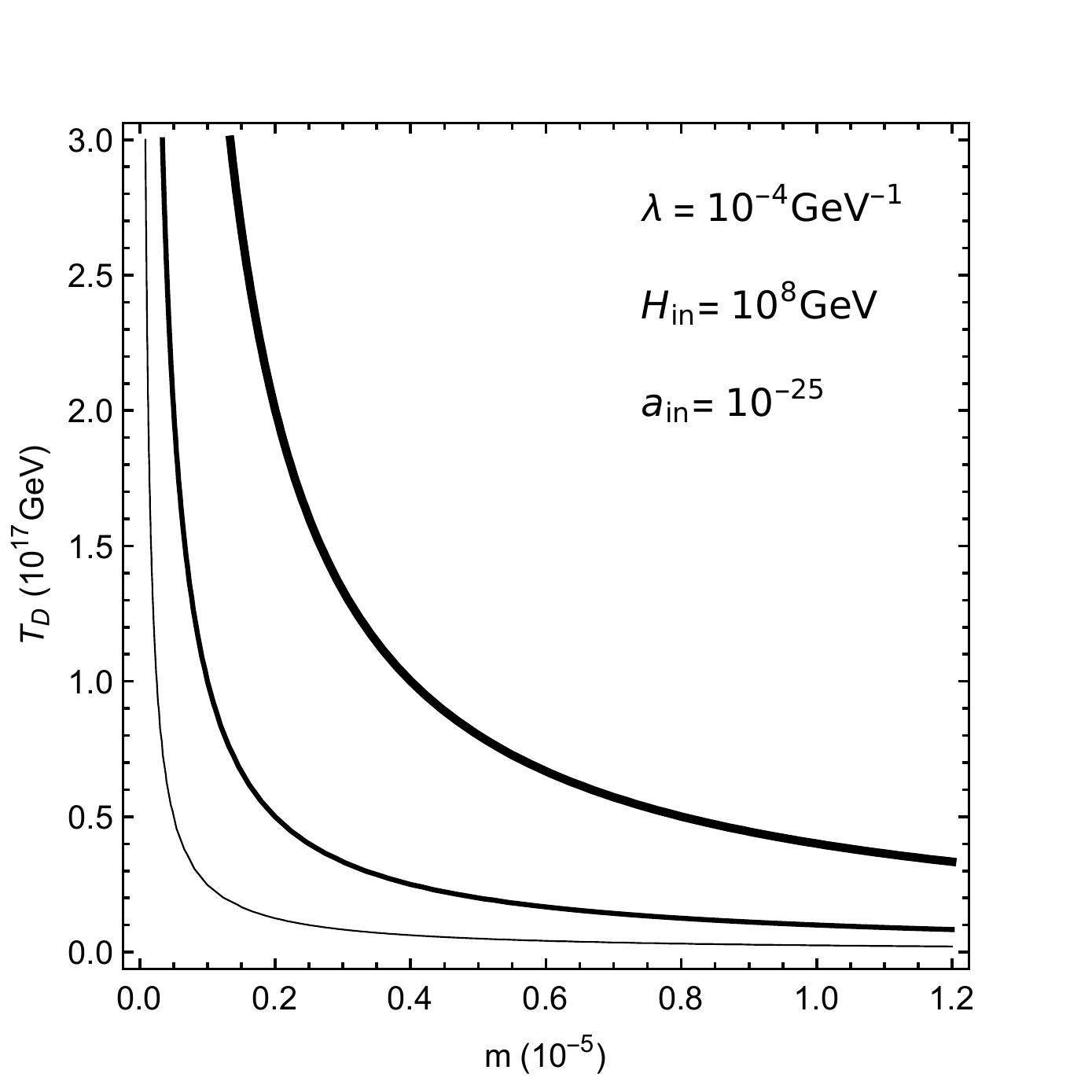}
    \centering
        \caption{The decoupling  temperature, $T_D$, as function of the~parameter $m$, in the model with varying $\alpha$ using the currently measured value of the asymmetry, $\eta_B~\simeq~8.6~\cdot~10^{-11}$.
  The thin line corresponds to  $\lambda_\alpha=~4~\cdot~10^{-19}$~GeV$^{-1}$, the middle line corresponds to  $\lambda_\alpha=2\cdot10^{-19}$ GeV$^{-1}$, and the thick line to $\lambda_\alpha= 10^{-19}$ GeV$^{-1}$.  The initial conditions are: $a_{in} = 10^{-25}$ and  $H_{in} = 10^{8}$ GeV.}
   \label{fig:alpha-TD_m}
\end{figure}

In order to calculate (\ref{BSBM_eta_B}), we need to solve the equation of motion of the field $\psi$ (\ref{BSBM_eom_1}).
 For the pure radiation $\mathcal{L}_{em}$ vanishes, so we can safely neglect this term. 
 Since we assumed, that the field $\psi$ is homogeneous and isotropic and only the null component of $J_B^\mu$ gives a contribution to the difference in the number densities, we can reformulate (\ref{BSBM_eom_1}) to:
\begin{align}
    \ddot{\psi} +3H\dot{\psi}= \frac{c^4}{16\pi G} \frac{\lambda_\alpha^2 c}{\Omega} \left( \frac{d}{dt}{J_B^0} +3H J_B^0 \right)\ . \label{BSBM_eom_2}
\end{align}
When we insert (\ref{Stat_Mech_current_T}), (\ref{Stat_Mech_baryon_excess_muT2}), and (\ref{BSBM_pot_chem}) into (\ref{BSBM_eom_2}), we can try to estimate the value of the right hand side of this equation and its impact onto the evolution of $\psi$ as:
\begin{align}
    \ddot{\psi}+ 3H\dot{\psi} = \frac{2\tilde{\beta}\dot{\psi} T \dot{T}}{ 1 -\tilde{\beta} T^2 }\ , \label{BSBM_eom_3} 
\end{align}
where $\tilde{\beta}$ is a constant of unit $K^{-2}$:
\begin{align} 
    \tilde{\beta}= \frac{g_i}{6\Omega} \frac{k_B ^2}{(\hbar c)^3} \left(\frac{c^4}{16 \pi G} \lambda_\alpha^2 \right)^2 \ . \quad 
\end{align}
 We assume that the temperature of the Universe in a short period of baryogenesis  did not change significantly  ($\dot{T} \approx 0$). We also evaluate the value of $\tilde{\beta} T^2$, which for $\lambda_\alpha=\lambda_{Pl}$ and $\omega \lambda$ of the order of few tens of MeV,  is much smaller than one.  For this reason we are excused to simplify (\ref{BSBM_eom_3}) to:
\begin{align}
    \ddot{\psi}+ 3H\dot{\psi}=0\ .\label{BSBM_eom_4}
\end{align}
 In order to solve (\ref{BSBM_eom_4}) we need to describe the dynamics of $\alpha$ by an explicit dependence on the scale factor $a(t)$.
 We make the following ansatz for $\alpha$: 
\begin{align}
    \alpha(t)= \alpha(t_0) \left[\frac{a(t)}{a(t_0)}\right]^m \label{BSBM_ansatz}\ ,
\end{align}
 where a constant parameter $m$ measures a change in $\alpha$ and the index ``$0$" denotes the current values of the fine structure constant, $\alpha(t_0)=\alpha_0 $, and the scale factor, $a(t_0)=a_0$. 
 A scenario with no variation of $\alpha$ can be restored for $m=0$. By inserting (\ref{BSBM_ansatz}) into (\ref{BSBM_eom_4}) we find that:
\begin{align}
    \frac{\ddot{a}(t)}{\dot{a}(t)}+2\frac{\dot{a}(t)}{a(t)}= 0\ . \label{BSBM_eom_final}
\end{align}
 The above equation can be integrated from $t$ to $t_{in}$, where time $t_{in}$ stands for the onset of baryogenesis. This gives  
\begin{align}
    a(t) =  a_{in} \left[ 1+3H_{in} (t-t_{in}) \right]^{1/3}.  \label{BSBM_scale factor}
\end{align}
 where $a(t_{in})=a_{in}$, and $H(t_{in})=H_{in}$ are, respectively, the initial value of the scale factor, and the corresponding Hubble parameter. 
The scale factor (\ref{BSBM_scale factor}) differs from the one which is expected for the radiation dominated Universe. The presence of the scalar field $\psi(t)$ shifts its value from  $a\sim t^{1/2}$  to $a\sim t^{1/3}$, this means that the solution (\ref{BSBM_scale factor}) scales  like a solution for the stiff fluid.
 
 By using (\ref{BSBM_psi1}), (\ref{BSBM_ansatz}) and (\ref{BSBM_scale factor}) we can find that
 
 \begin{figure*}[ht!]
  \includegraphics[width=8.5cm]{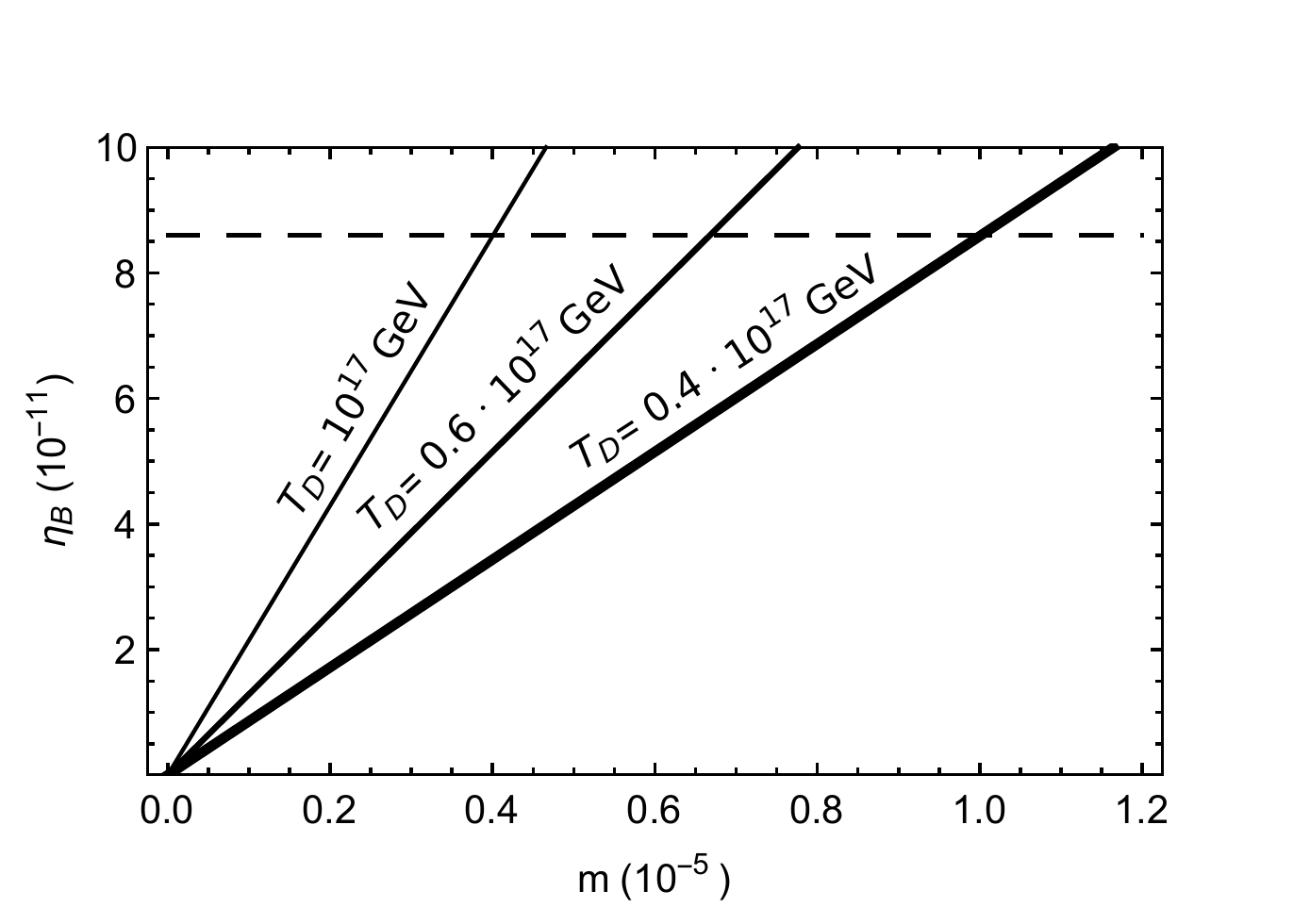}
    \includegraphics[width=8.5cm]{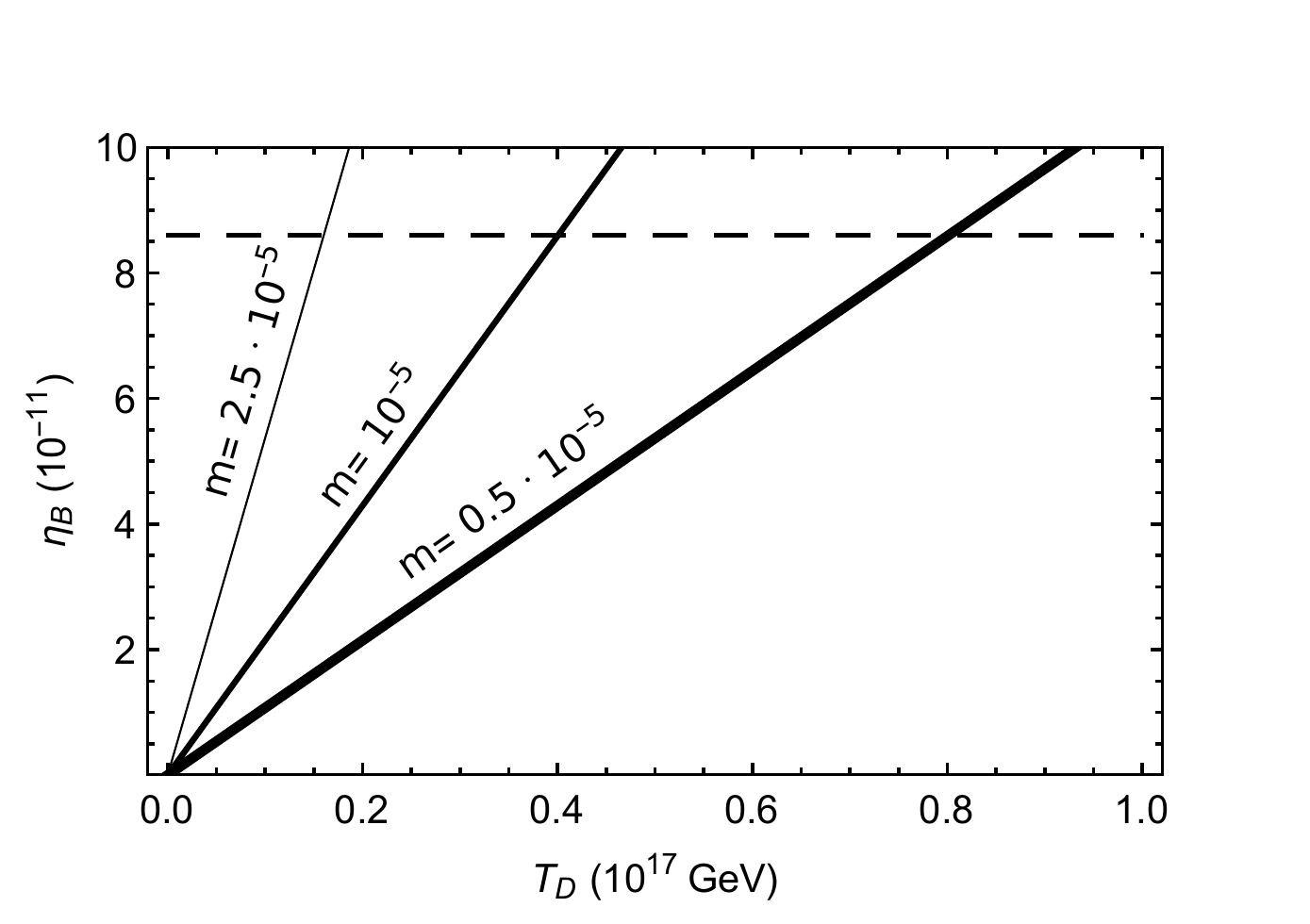}
 \caption{The baryon to entropy ratio $\eta_B$ (\ref{BSBM_final_assym}) as a function of the parameter $m$ (left) and the decoupling temperature $T_D$ (right) in the varying $\alpha$ model. 
 The horizontal dashed lines indicate the currently measured value of the asymmetry, $\eta_B\simeq 8.6\cdot10^{-11}$. 
 The plot on the left was made for the three values of the decoupling temperature:  $T_D=10^{17}$  GeV (thin left line), $T_D=0.6\cdot10^{17}$ GeV (middle line) and $T_D=0.4\cdot10^{17}$ GeV (thick right line). 
 The plot on the right was made for three values of the parameter $m$:  $m=2.5\cdot10^{-5}$ (thin left line), $m=10^{-5}$ (middle line), and $m=0.5\cdot10^{-5}$ (thick right line). 
 All plots were made for: $a_{in} = 10^{-25}$, $H_{in} = 10^{8}$ GeV,  $\lambda = 10^{-4}$ GeV$^{-1}$, and $\lambda_\alpha = 10^{-19}$ GeV$^{-1}$. }\label{fig:alpha-eta_m_TD}
\end{figure*}

 
 \begin{align}
     \psi(t)= \frac{1}{2}\ln \left|  \left[ 1+3H_{in}(t-t_{in}) \right]^{\frac{m}{3}}\left(\frac{a_{in}}{a_0}\right)^m \right| \label{BSBM_psi_sol}
 \end{align}
 and 
 \begin{align}
     \dot{\psi}(t)= \frac{1}{2} m H_{in} \left[ 1+3H_{in}(t-t_{in}) \right]^{-1} .
 \end{align}
In the limit $m \to 0$ the solutions (\ref{BSBM_scale factor}) and (\ref{BSBM_psi_sol}) solve simultaneously all the field equations when:
\begin{align}
    \rho_{em0}=\left(\frac{3}{8\pi G}- \frac{\Omega m^2}{8 c^4 } \right) \left(H_{in}\frac{6}{6+m} \right)^2 \left(\frac{a_{in}}{a_0}\right)^{6+m} \ ,
\end{align}
 for $\rho_{em0}$ being a positive constant in the expression for the energy density of the stiff fluid:
 \begin{align}
     \rho_{em}=\rho_{em0} \left( \frac{a_0}{a}\right)^6\ .
 \end{align}
In order to write $\eta_B$ as a function of the temperature~$T$, we combine the Friedmann equation (\ref{BSBM_Friedmann_eq}) and the energy density (\ref{Stat_Mech_entropy_density}) to  yield 
\begin{align}
   &\left[ 1+3H_{in}(t-t_{in})\right]^{\frac{6-m}{3}}= \\ 
   &\left( \frac{3}{8 \pi G}- \frac{\Omega m^2}{8c^4}\right)\left[ \frac{\pi ^2 g_i}{30c^2} \frac{k_B ^4}{(\hbar c)^3} \left(\frac{a_0}{a_{in}}\right)^m \right]^{-1} H_{in}^2 T^{-4}\ , \nonumber
\end{align}
and finally express (\ref{BSBM_eta_B}) in terms of temperature as:
\begin{align}
\label{BSBM_final_assym}
    &\eta_B = \frac{1}{2} m H_{in} \frac{15c^3}{16\pi G} \frac{\lambda^2_\alpha g_i}{8\pi^2 g_{*s}} \frac{1}{k_B^2 T} \left\{H_{in}^2 T^{-4} \times \right. \\
   &\left. \left(\frac{3}{8 \pi G}- \frac{\Omega m^2}{8c^4}\right) \left[ \frac{\pi ^2 g_i}{30 c^2} \frac{k_B ^4}{(\hbar c)^3} \left(\frac{a_0}{a_{in}}\right)^m \right]^{-1} \right\}^{-\frac{3}{6-m}} \nonumber. 
\end{align}

We have found that it is possible to achieve the currently measured value of the baryon asymmetry, $\eta_B$, in the BSBM model of baryogenesis, as well as in the model of varying $G$ discussed in \ref{G_chapter}. A possible parameter space is presented in Fig. \ref{fig:alpha-TD_m} for three different values of $\lambda_\alpha$. We have compared the ansatz (\ref{BSBM_ansatz}) with the measurements of time variation of $\alpha$ to find the order of magnitude of the parameter $m$. We have found $m$ to be in the range:
\begin{equation}
    m= -\log_{1+z_\alpha}\left(1+\frac{\Delta \alpha}{\alpha}\right), 
 \end{equation}
 where $z_\alpha$ is a corresponding value of redshift for which $\Delta \alpha/\alpha$ has been measured.
 Using the bound from Ref. \cite{alphaobs} we have decided to restrict $m$ to be of the order of $10^{-6}$. 
 This corresponds to the decoupling temperature $T_D~\sim~10^{16}$ GeV. 
 In our model $m$ takes positive values only, which stands for the smaller $\alpha$ in the past. 
 However, positive $m$ can also be admitted according to the so-called $\alpha$--dipole measurement \cite{alphaobs}. 
 Similarly to the model of $G$-driven baryogenesis, any small change in the initial value of the scale factor $a(t)$ at the moment of baryogenesis does not have any strong impact on $\eta_B$. 
 The sensitivity of $\eta_B$ increases with the growth of $m$. 
 Again, the second most significant parameter is the length $\lambda_\alpha$, which was chosen to be of the order of $10^{-19}$~GeV (see Fig. \ref{fig:alpha-TD_m}). 
 The baryon to entropy ratio $\eta_B$ (\ref{BSBM_final_assym}) as a function of the parameter $m$ for three values of the decoupling temperature $T_D$ (see the plot on the left), as well as a function of the temperature $T_D$ for three values of $m$ (see the plot on the right) has been shown in Fig. \ref{fig:alpha-eta_m_TD}. 

\subsection{Dynamical speed of light $c$ baryogenesis}
\label{c_chapter}


Early ideas about varying speed of light $c$ were even distributed by Einstein \cite{Einstein1907} and then many years later recalled by Petit \cite{Petit1988} and Moffat \citep{Moffat93A,Moffat93B}. Moffat developed a fully consistent theory which was designed to alternatively solve all the problems of standard cosmology which were originally resolved by the inflationary scenario \cite{Guth81}. Different types of varying speed of light models were also suggested by Albrecht and Magueijo \cite{Albrecht99}, Barrow and Magueijo \cite{BM99,Barrow99}, and further developed by Magueijo \cite{Mag2001,Magueijo00}.  These models are also useful to solve the standard cosmological problems such as the horizon problem, the flatness problem, the $\Lambda-$problem, and has recently been proposed to solve the singularity problem \cite{JCAP13}. Another different class of varying speed of light models was given by Avelino and Martins \cite{AM}. All the above models have recently been subject to statistical evaluation against observational data \cite{ApJ17} showing the preference of Moffat's models, which we have selected to study in the context of baryogenesis.  

Here we combine the most recent Moffat's approach \cite{Moffat16} with the theory of the spontaneous baryogenesis. The appropriate action is made up of three terms:
\begin{equation}
    S= S_{\Phi} + S_{m} + S_B\ , \label{c_action}
\end{equation}
where
\begin{align}
    S_{\Phi}&=  \frac{1}{16 \pi G}\int dx^4 \sqrt{-g} \left( \Phi R - \frac{\kappa}{\Phi}\partial_\mu \Phi \partial^\mu\Phi\right)\ ,\label{c_S_Phi} \\
    S_{m} &=\int dx^4  \sqrt{-g} \mathcal{L}_{m} \label{c_S_m}\ , \\
    S_B &= \frac{1}{16 \pi G}\int dx^4 \lambda_c^2 (\partial_\mu\Phi) J_B^\mu \label{c_S_B}\ .
\end{align}
The action (\ref{c_S_Phi}) is the gravitational action with the field $\Phi$ coupled to the curvature and the kinetic term with a constant $\kappa$. We also introduce the matter term $S_{m}$, since at the moment of baryogenesis the Universe was filled-in  with radiation. The baryon asymmetry is produced by the interaction term (\ref{c_S_B}). As in the previous chapters, the length $\lambda_c$ is the cut-off length of the applicability of the theory, $J_B^\mu$ is the baryon current, and its null component describes a difference in the particle and the antiparticle number densities (cf. eq. (\ref{Stat_Mech_current_T})). 
Similarly to the section \ref{G_chapter}, we follow the notation of Refs. \cite{LL,Will,JCAP15} for the Einstein--Hilbert action.

This means that we take $x^\nu= (x^0, x^1, x^2, x^3)$ and so (\ref{FRW}) is replaced by 
\begin{equation}
    ds^2=-(dx^0)^2+ a^2(x^0)\left[ \frac{dr^2}{1-kr^2}+r^2 d\theta^2+r^2 \sin^2 \theta d\tilde{\phi}^2 \right],
    \label{c_metric}
\end{equation}
where $x^0=c(t)t$ (cf. the discussion of Appendix A in Ref. \cite{Albrecht99}). As a consequence, the dynamics of the speed of light field $\Phi$ is given by:
\begin{equation}
    \Phi(x^\nu)= c^3(x^\nu)\ , \label{c_field}
\end{equation}
which differs from the Moffat's definition of the $\Phi$ field in \cite{Moffat93A} and also in other references which take $\Phi=c^4$ \cite{Albrecht99,Barrow99}. However, both formulations are equivalent. 

In fact, the original Moffat's theory consists of the action representing the dynamics of four scalar fields. In a later paper \cite{Moffat16} a vector field was driving the spontaneous violation of $SO(3,1)$ Lorentz invariance, while a dimensionless scalar field, minimally coupled to gravity, was responsible for quantum primordial fluctuations. Nevertheless, unlike in \cite{Moffat93A}, we are dealing with a small Lorentz violation, and consequently with a small change in the speed of light $c$. In this paper we do not intend to explain the fast exponential expansion of the early universe to make it alternative to inflation and for this reason we have dropped the part of the Moffat's theory, which exhibits the strong Lorentz symmetry breaking and therefore, the large change in $c$ ($c\approx10^{28}c_0$, where $c_0$ is the current value of the speed of light). The Lagrangian for the quantum primordial fluctuations has not been included, either. At the moment of baryon asymmetry generation, this term is not relevant anymore, and can safely be neglected.  

The variation of (\ref{c_action}) with respect to the metric $g^{\mu\nu}$ leads to the field equations:
\begin{align}
    G_\mu^\nu= \frac{8\pi G}{\Phi^{4/3}} \left( 
    {T_m}_\mu^{~\nu} + 
    {T_\Phi}_\mu^{~\nu} + 
    {T_B}_\mu^{~\nu}
    \right)\ , \label{c_G_mu^nu}
\end{align}
where the tensors ${T_i}_\mu^{~\nu}$ are given by:
\begin{align}
    {T_m}_{\mu}^{~\nu} &= g_{\mu\sigma}\frac{2}{\sqrt{-g}}\frac{\partial}{\partial  g_{\sigma\nu}}\left(
    \sqrt{-g}\mathcal{L}_{m}\right)  \label{c_Tm}          \ , \\
    {T_\Phi}_\mu^{~\nu} &= \frac{\Phi^{4/3}}{8 \pi G}\left(\frac{1}{\Phi} \nabla_{\mu}\nabla^{\nu}\Phi - \delta_{\mu}^{\nu}\Box \Phi \right)\nonumber \label{c_T_Phi} \\
    &+ \frac{\Phi^{4/3}}{8 \pi G}\frac{\kappa}{\Phi^2}\left( \partial_{\mu} \Phi \partial^{\nu} \Phi - \frac{1}{2} \delta_{\mu}^{\nu} \partial_{\beta} \Phi \partial^{\beta} \Phi \right)\ ,\\
    {T_B}_\mu^{~\nu} &= \frac{\Phi^{1/3}}{16\pi G} \lambda_c^2 \delta_\mu^\nu \partial_\gamma\Phi J_B^\gamma \label{c_TB}\ .
\end{align}

The equation of motion of the field $\Phi$ takes the form:
\begin{align}
    \Box \Phi &= \frac{8 \pi G}{(3+2\kappa)\Phi^{1/3}} T_{m}\label{c_eq.of.motion_1} \\
   & +\frac{\lambda_c^2}{3+2 \kappa} \Phi \left( \partial_\mu J_B^\mu + J_B^\gamma {\Gamma^\mu}_{\mu \gamma } + \frac{2}{\Phi} \partial_\mu \Phi J_B^\mu     \right)\nonumber\ ,
\end{align}
where ${\Gamma^\mu}_{\mu \gamma }$ are the Christoffel connection coefficients, and $T_m$ is the trace of the radiation energy--momentum tensor. We assume the barothropic equation of state of the fluid $p~=~w \rho c^2$, which gives the trace $T_m= -\rho c^2(1-3w)$ and vanishes for pure radiation field, $w=1/3$. The equation (\ref{c_eq.of.motion_1}) reads then as:
\begin{align}
    \Phi '' + 3\tilde{H}\Phi ' = - \frac{\lambda_c^2}{3+2 \kappa} \Phi \left(  {J_B^0} '+ 3\tilde{H}J_B^0  + 2\tilde{H}_\Phi {J_B^0} '     \right) \label{c_eom2}\ ,
\end{align}
where ($'$) stands for the derivative with respect to the coordinate $x^0$, $\tilde{H} = a'/a$ is the Hubble parameter, and $\tilde{H}_\Phi=\Phi'/\Phi$. Both, $\tilde{H}$ and $\tilde{H}_\Phi$ are of the unit $m^{-2}$, instead the usual $s^{-2}$. This is a consequence of a chosen definition of the action (\ref{c_action}) and of coordinates in (\ref{c_metric}). 
The spontaneous baryogenesis occurs when CPT symmetry is broken in the Universe, which is in thermal equilibrium. 
This leads to a conclusion that particles, as well as the antiparticles are in thermodynamical equilibrium, but possess different energies. This is what we call the energy shift and can find it by investigating the contribution of (\ref{c_S_B}) to the total energy density. The chemical potential takes the form: 
\begin{align}
    \mu_B=E_B-E_{\bar{B}}= \frac{\lambda_c}{16 \pi G}\Phi^{1/3} \Phi' \label{c_chem_pot} .
\end{align}
Unlike in the previous sections \ref{G_chapter} and \ref{al_chapter} of varying $G$ and varying $\alpha$, here not only the derivative of the field $\Phi'$ enters the chemical potential $\mu_B$, but also the field $\Phi$ itself. The ratio of the baryon asymmetry to the entropy density is given by:
\begin{align}
    \eta_B= \frac{15 g_i}{4 \pi^2 g_{*s}} \frac{1}{k_B^2 T} \mu_B \label{c_etaB1} \ ,
\end{align}
which together with (\ref{c_chem_pot}) yields:
\begin{align}
\eta_B= \frac{\lambda_c^2}{16\pi G}\frac{15 g_i}{4 \pi^2g_{*s}}\frac{1}{k_B^2 T} \Phi^{1/3}\Phi' \label{c_etaB2}\ .
\end{align}
In order to find the value of (\ref{c_etaB2}), we need to solve the equation of motion (\ref{c_eom2}). 
First, we find $J_B^0$, which is the matter--antimatter excess:
\begin{align}
    J_B^0= \frac{g_i k_B^2}{6\hbar^3}\frac{\lambda_c^2}{16 \pi G}\Phi^{-2/3}\Phi' T^2 \label{c_J0} 
\end{align}
and then insert (\ref{c_J0}) into (\ref{c_eom2}). This gives a relation, which connects $\Phi$ and its derivatives with the temperature $T$:
\begin{align}
    \Phi '' +3\tilde{H} \Phi' = \frac{2\chi \left( \frac{2}{3} \Phi^{-2/3}\Phi^{'2} T^2 + \Phi^{1/3} \Phi'T T'\right)}{1-\chi \Phi^{1/3} T^2}\ , {\label{c_eom3}}
\end{align}
where $\chi$ is a constant of the unit $K^2 m s^{-1}$:
\begin{align}
  \chi= -  \frac{g_i k_B^2}{6\hbar^3}\frac{\lambda_c^4}{16 \pi G (3+2\kappa)}\ .
\end{align}
It has been checked, that the right hand side of (\ref{c_eom3}) is small, and so it can safely be neglected. A resulting simplified equation of motion is then:
\begin{align}
     \Phi '' +3\tilde{H} \Phi' =0 \label{c_eom4}\ .
\end{align}
In order to solve (\ref{c_eom4}), we make an ansatz for the field $\Phi$ as follows:
\begin{align}
    \Phi(x^0)= c^3_0\left[\frac{a(x^0)}{a(x^0_0)}\right]^{3n} \ . \label{c_Phi}
\end{align}
where $x_0^0= c_0t_0$ and $n$ is a parameter, which indicates the variation in $c$ (presumably small since we deal with approximate Lorentz symmetry). In the limit $n \to 0$, a currently measured value of the speed o light $c_0$ is restored and the field is just equal to $\Phi=c_0^3$. We will denote $a(x^0)\equiv a$ and $a(x^0_0)\equiv a_0$, later on.
The solution of (\ref{c_eom4}) is: 
\begin{align}
    \Phi = \Phi_{in} \left[1+3\tilde{H}_{in}(n+1)\left(x^0-x^0_{in}\right) \right]^{\frac{n}{n+1}} \ , \label{c_Phi_solution}
\end{align}
where $\Phi_{in}=c_0^3 (a_{in}/a_0)^{3n}$ is the initial value of the field at the beginning of baryogenesis, $x_{in}^0= c_{in}t_{in}$, , and $\tilde{H}$ is the initial value of the Hubble parameter. 
The solution (\ref{c_Phi_solution}) solve the full set of  the field equations in the limit $n \to 0$ when: 
\begin{align}
    \tilde{\rho}_{m0}=
    \frac{3 \Phi_{in}^{2/3}}{8 \pi G}
    \left(\frac{a_{in}}{a_0}\right)^{6}
    \left(3H_{in}\frac{n+1}{n+3} \right)^2 
    \left( 1+3n-\frac{3\kappa}{2}n^2\right)\ ,
\end{align}
 for $\tilde{\rho}_{m0}$ being a positive constant in the expression for the energy density of the stiff fluid:
 \begin{align}
     \tilde{\rho}_{m}=\tilde{\rho}_{m0} \left( \frac{a_0}{a}\right)^6\ .
 \end{align}
Taking the derivative of (\ref{c_Phi_solution}), we find the final expression for the baryon asymmetry $\eta_B$:
\begin{align}
    \eta_B =& \frac{\lambda_c^2}{16\pi G}\frac{15 g_i}{4 \pi^2g_{*s}}\frac{1}{k_B^2 T} \times \label{c_etaB30} \\
    &3n\tilde{H}_{in} \Phi^{4/3}_{in} \left[1+3\tilde{H}_{in}(n+1)\left(x^0-x^0_{in}\right) \right]^{\frac{\frac{1}{3}n-1}{n+1}} 
    \ . \nonumber
\end{align}

\begin{figure}
  \includegraphics[width=8.5cm]{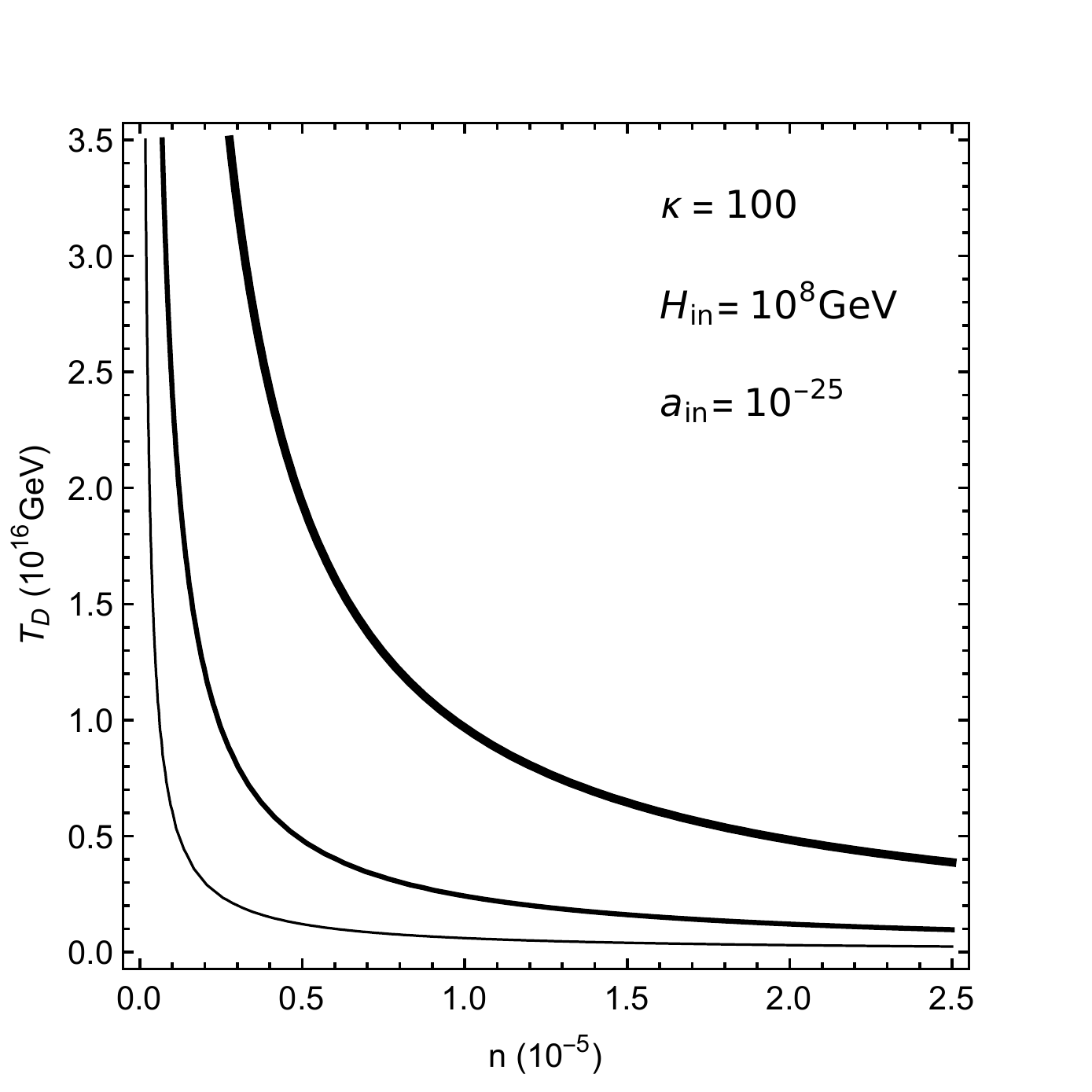}
  \caption{The decoupling  temperature, $T_D$, as function of the~parameter $n$, in the model with varying $c$ using the currently measured value of the asymmetry, $\eta_B~\simeq~8.6~\cdot~10^{-11}$, and for $\kappa=100$. 
 The thin line corresponds to  $\lambda_c= 4\cdot10^{-19}$~GeV$^{-1}$, the middle line corresponds to  $\lambda_c=2\cdot10^{-19}$ GeV$^{-1}$, and the thick line to $\lambda_c= 10^{-19}$ GeV$^{-1}$.  The initial conditions are: $a_{in} = 10^{-25}$ and  $H_{in} = 10^{8}$ GeV.}  \label{fig:c-Td-n}
\end{figure}
The relation (\ref{c_etaB30}) is a function of the temperature $T$ and the $x^0$--coordinate. In order to express $\eta_B$ as a function of temperature only, we use the temperature dependent expression for the energy density of relativistic particles (\ref{Stat_Mech_en_density_temp}), which depends on temperature, as well as on the speed of light $c$ at a given moment. For this reason, we find it reasonable to replace $c^3$ in the denominator by the field $\Phi$. By combining the modified equation (\ref{Stat_Mech_en_density_temp}) with the Friedmann equation for a flat universe:
\begin{align}
    \tilde{H}^2= \frac{8\pi G}{3} \Phi^{-2/3}\left( \rho_m+\rho_\Phi+\rho_B\right)\ , \label{c_Fied_eq}
\end{align}
where: 
\begin{align}
    \varepsilon_m=& \rho_m\Phi^{2/3}= g_*\frac{\pi^2}{30} \frac{ k_B^4}{ \hbar^3\Phi}T \cdot \Phi^{2/3}\ ,  \\ 
    \varepsilon_\Phi=& \rho_\Phi\Phi^{2/3}=\frac{\Phi^{4/3}}{8 \pi G} \left(  \frac{\kappa}{2} \tilde{H}_\Phi^2 - 3\tilde{H}\tilde{H}_\Phi\right) \ ,\\  
    \varepsilon_B=& \rho_B\Phi^{2/3}= \frac{\Phi^{4/3}}{16 \pi G} \lambda_c \tilde{H}_\Phi J_B^0\ , 
\end{align}
we can find an approximate time--temperature relation in the varying $c$ models for the radiation dominated epoch. In order to do so, we have neglected $\varepsilon_B$ in (\ref{c_Fied_eq}) treating this term as a small perturbation on the background of the main fluid, which is radiation (similarly as we did in (\ref{c_eom4})). This leads to:
\begin{align}
    &\left[1+3\tilde{H}_{in}(n+1)\left(x^0-x^0_{in}\right)\right]^{-1/3} =  \label{c_t-T relation}\\
    &\left[\Phi_{in}^{7/3} H_{in}^2 \left(\frac{90\hbar^3g_*^{-1}}{8 \pi^3 G k_B^4}\right)\left(-\frac{3\kappa}{2}  n^2+ 3n+1  \right)   T^{-4}\right]^{ \frac{n+1}{n-6}}\nonumber \ .
\end{align}
The limits for the parameter n are as follows 
\begin{equation}
   n \in \left(\frac{3-\sqrt{9+6\kappa}}{3\kappa},    \frac{3+\sqrt{9+6\kappa}}{3\kappa}\right)\  ,
\end{equation}
for $\kappa > 0$, and 
\begin{equation}
   n \in  \left( -\infty, \frac{3+\sqrt{9+6\kappa}}{3\kappa}\right) \cup \left(\frac{3-\sqrt{9+6\kappa}}{3\kappa},+\infty\right)    \ ,
\end{equation}
for $\kappa<0$. For $\kappa=0$, $n> -1/3$. Finally, we find that the baryon asymmetry reads as 
\begin{align}
    \eta_B& = \frac{\lambda_c^2}{16\pi G}\frac{15 g_i}{4 \pi^2g_{*s}}\frac{1}{k_B^2 T} 3n\tilde{H}_{in} \Phi^{4/3}_{in}  \times  \label{c_etaB3} \\
    &
    \left[\Phi_{in}^{7/3} H_{in}^2  \left(\frac{90\hbar^3g_*^{-1}}{8 \pi^3 G k_B^4}\right) \left(-\frac{3\kappa}{2}  n^2+ 3n+1  \right) T^{-4}\right]^{ \frac{3-n}{n-6}}. \nonumber
    \label{c_etaB3}
\end{align}

\begin{figure*}[ht!]
  \includegraphics[width=8.5cm]{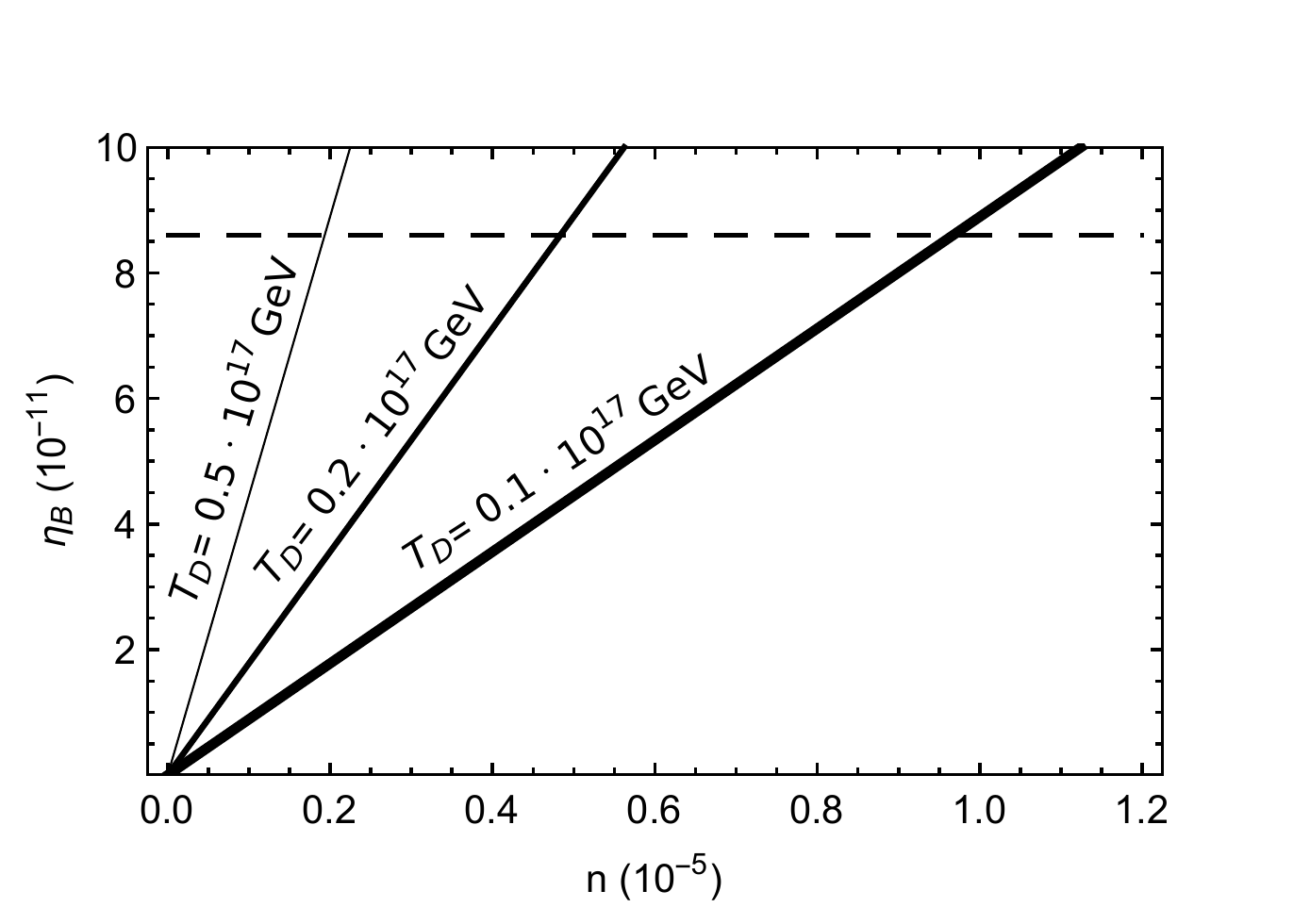}
    \includegraphics[width=8.5cm]{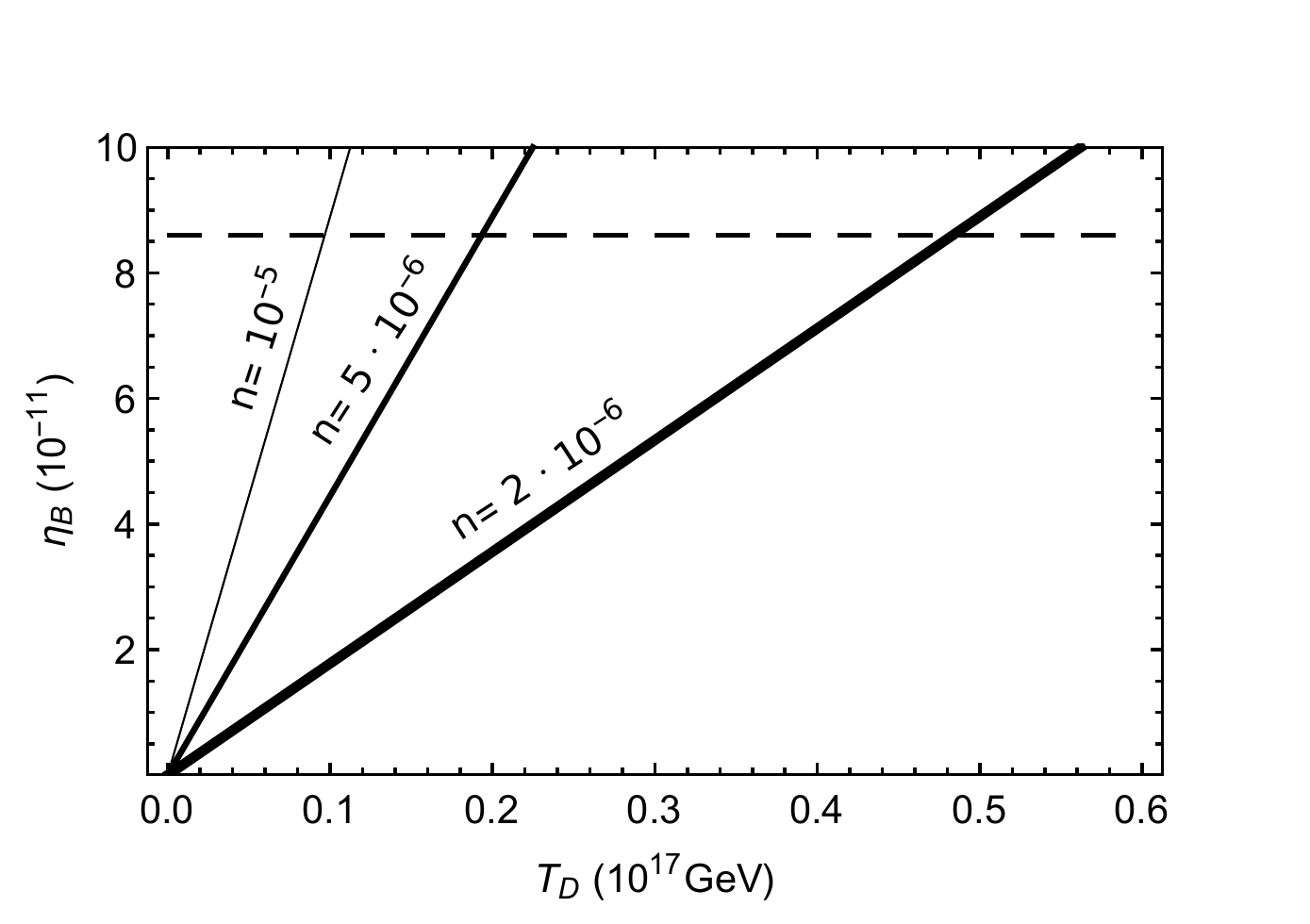}
 \caption{The baryon to entropy ratio $\eta_B$ (\ref{c_etaB3}) as a function of the parameter $n$ (left) and the decoupling temperature $T_D$ (right) in the varying $c$ model.  The dashed lines indicate the currently measured value of the asymmetry, $\eta_B\simeq 8.6\cdot10^{-11}$.  The plot on the left was made for the three values of the decoupling temperature:  $T_D=0.5\cdot10^{17}$  GeV (thin left line), $T_D=0.2\cdot10^{17}$ GeV (middle line) and $T_D=0.1\cdot10^{17}$ GeV (thick right line).  The plot on the right was made for the three values of the parameter $n$:  $n=10^{-5}$ (thin left line), $n=5 \cdot10^{-6}$ (middle line), and $n=2\cdot10^{-6}$ (thick right). All plots were made for: $a_{in} = 10^{-25}$, $H_{in} = 10^{8}$ GeV, $\lambda_c = 10^{-19}$ GeV$^{-1}$, and $\kappa=100$. }\label{fig:c-eta_n_TD}
\end{figure*}
 In summary, for the Moffat's model of the varying speed of light, $c$, we have found the parameter space, for which the desirable asymmetry is possible for a given range of the parameter $n$ and the temperature $T_D$ (see the Fig. \ref{fig:c-Td-n}). We have found a relation between $n$ and the constant $\kappa$ from  (\ref{c_t-T relation}). The limits on the $n$ values has been shown in the Table \ref{tab:c-n_limits}. We have noticed, that the limits corresponding to the negative and positive $\kappa$ partially overlap and have decided to proceed the calculation only for positive $\kappa$, even though there are no observational or experimental bounds on its value. Nevertheless, for the chosen order of magnitude of the parameter $n$, the influence of $\kappa$ onto the final result is negligible. However, $\kappa$ becomes more relevant, when bigger $n$ is taken into account. In order to estimate an order of magnitude of $n$, we have assumed that any possible $c$ variation would find its manifestation in the variation of the fine structure constant and thus it would become visible in the measurement of the $\alpha$--time variation. By comparison of the value of $\Delta\alpha/\alpha$ with the ansatz (\ref{BSBM_ansatz}) we have achieved the following  expression for $n$: 
\begin{equation}
    n= \log_{1+z_\alpha}\left(1+\frac{\Delta \alpha}{\alpha}\right).
 \end{equation}
Calculated limits on the parameter $n$ have turned to be of the order of $\sim 10^{-6}$. The corresponding temperature for baryogenesis is about $10^{16}$ GeV. However, the value of $n$ does not need to be necessarily compared with the results for time variation of $\alpha$, and one could consider even a bigger change of the speed of light.  Our model favours the positive values of $n$, and thus the increase of the speed of light. This is not in the spirit of the varying speed of light models, which solve the basic cosmological problems and stand as an alternative to the inflation theories. In Fig.~\ref{fig:c-eta_n_TD} we have shown the baryon to entropy ratio $\eta_B$ (\ref{c_etaB3}) as a function of the parameter $n$ for three possible values of the temperature $T_D$ (see the plot on the left) and as a function of the temperature $T_D$ for three possible values of $n$ (see the plot on the right).

\begin{table}[b!]
\centering
\caption{Limits for the parameter $n$ for some specific values of   $\kappa$}
\label{tab:c-n_limits}
\begin{tabular}{c|c}
\hline\hline
 \rule{0pt}{15pt}$\kappa$ & $n$ \\[7pt] \hline\hline
\rule{0pt}{10pt}
-1    &   (-$\infty$, -1.5774) $\cup$  (-0.4227, + $\infty$) \\[3pt]
-$\frac{1}{2}$    & (-$\infty$, -3.6330) $\cup$ (-0.3670, + $\infty$)                    \\[3pt]
0    & (-1/3, +$\infty$)                     \\[3pt]
1   & (-0.2910, 2.2910)                     \\[3pt]
10   & (-0.1769, 0.3769)                    \\ [3pt]
100   & (-0.0723, 0.0923)                     \\[3pt]
1 000  & (-0.0248, 0.0268)                     \\[3pt]
10 000   & (-0.0081, 0.0083)                     \\[3pt]
$10^{10}$  & (-0.0248, 0.0242)                     \\[3pt]
$\infty$      &              0                                    \\[3pt]
\hline
\end{tabular}
\end{table}

\section{Results and Conclusions}
\label{discuss}

In this paper we have investigated the scalar fields for the dynamical constants: the gravitational constant $G$, the fine structure constant $\alpha$, and the speed of light $c$, which, as we have assumed, could drive the baryogenesis in the universe. The spontaneous baryogenesis model was investigated, in which  the baryon number violating processes occur in thermal equilibrium, while the Universe allows a period of CPT symmetry breaking. We have formulated and solved the dynamical equations for the scalar fields corresponding to varying $G$, $\alpha$, and $c$, acting as thermions. We have applied some special ans\"{a}tze for the scale factor for each of the fields, as given in (\ref{BD_G_ansatz}), (\ref{BSBM_ansatz}), and (\ref{c_Phi}), which related the scalar fields with the evolution of the scale factor, and the resulting parameters $q$, $m$, and $n$ which measured the degree of variability of $G$, $\alpha$, and $c$, accordingly.  We have calculated the cosmological equations and used them to find the relation between the time $t$ and the temperature $T$ in the radiation dominated epoch. We have used the standard statistical mechanics tools in order to introduce the temperature dependence into the fields $\phi(G(x^0))$, $\psi(\alpha(x^0))$, $\Phi(c(x^0))$ (where the coordinate $x^0 = ct$), and in order to calculate the baryon asymmetry ratio $\eta_B$ of the net number density of baryons and antibaryons to the entropy density of photons. 

As a result  of our calculations, we have obtained similar conclusion as in the previous literature i.e. that varying $G$ can drive baryogenesis in the universe. Our main new results (not yet considered in the literature) are obtained for varying fine structure constant $\alpha$ models, as well as for varying speed of light $c$ models. We have shown that in each of these frameworks the current observational value of the baryon to entropy ratio $\eta_B \sim 8.6 \times 10^{-11}$ can be obtained for large set of parameters $q, m, n$, as well as the decoupling temperature $T_D$, and the characteristic cut-off length scale $\lambda$. This means that not only varying-$G$-driven baryogenesis is possible, but also varying-$\alpha$-driven and varying-$c$-driven baryogeneses are admissible. 

It is advisable to note that there exist models in which two of the three considered in this paper constants vary simultaneously \cite{Barrow99,varGc,adam2015,varGalpha}. However, in the most interesting case of varying $G$ and $\alpha$ it has been shown that an overall evolution of the universe is determined by $G$ and follows Brans-Dicke model so an extra influence of $\alpha$ on baryogenesis is not expected to be large. It is then expected that similar small effect of $c$-variability would remain in both varying $G$ and $c$ models. The detailed quantitative considerations of such models will be considered in some future work.

\section{Acknowledgements}

This project was financed by the Polish National Science Center Grant DEC-2012/06/A/ST2/00395. M.P.D. wishes to thank Alan Kosteleck\'y, Ralf Lehnert, Joao Magueijo, John Moffat, Eray Sabancilar, and John Webb for discussions. K.L. wishes to thank to Vincenzo Salzano and Tom\'{a}\v{s} Husek for discussions and valuable suggestions.

\end{document}